\documentclass{optica-article}

\journal{opticajournal}
\articletype{Research Article}
\titleformat{\section}{\normalsize\bfseries}{}{0em}{}
\titleformat{\subsection}{\normalsize\itshape}{}{0em}{}

\usepackage{amsmath,amssymb,mathtools}
\usepackage{bm}
\usepackage{booktabs}
\usepackage{float}
\usepackage[detect-all]{siunitx}
\usepackage{caption,subcaption}
\usepackage{microtype}
\usepackage{tikz}
\usetikzlibrary{arrows.meta,positioning,calc}
\usepackage{xcolor}

\newcommand{\Vscmax}{V_\mathrm{sc}}
\newcommand{\Vfb}{V_\mathrm{fb}}
\newcommand{\Vpv}{V_\mathrm{pv0}}
\newcommand{\VT}{V_\mathrm{tr}}
\newcommand{\Veff}{V_\mathrm{eff}}
\newcommand{\Vapp}{V_\mathrm{app}}
\newcommand{\epss}{\varepsilon_\mathrm{scr}}
\newcommand{\Df}{\Delta{}s}
\newcommand{\Pcav}{P_\mathrm{eff}}
\newcommand{\gzero}{g_0}
\newcommand{\krel}{k_\mathrm{rel}}
\newcommand{\Rtrap}{R}
\newcommand{\rV}{r_V}
\newcommand{\fp}{s_+}
\newcommand{\fm}{s_-}
\newcommand{\Peclet}{\mathrm{Pe}}

\begin{document}

\title{Voltage-Enhanced Photorefractive Hysteresis in Thin-Film
  Lithium Niobate Photonic-Crystal Cavities}

\author{Alessandro Buzzi,\authormark{1,*}
  Bjarke~A. Frederiksen,\authormark{1,2}
  Mikkel Heuck,\authormark{2}
  Dirk~R. Englund\authormark{1}
  and Dashiell~L.\,P. Vitullo,\authormark{1,3,\textdagger}}

\address{%
  \authormark{1}Research Laboratory of Electronics,
  Massachusetts Institute of Technology,
  Cambridge, Massachusetts 02139, USA\\
  \authormark{2}Department of Electrical and Photonics Engineering,
  Technical University of Denmark,
  Kgs.\ Lyngby 2800, Denmark\\
  \authormark{3}DEVCOM Army Research Laboratory,
  Adelphi, Maryland 20783, USA}

\email{\authormark{*}abuzzi@mit.edu, \authormark{\textdagger}dashiell.l.vitullo.civ@army.mil}

\maketitle

\begin{abstract}
We observe voltage-enhanced photorefractive hysteresis in thin-film
lithium niobate nanocavities, causing the resonance to appear at different
wavelengths during increasing- and decreasing-wavelength laser scans.
Their separation grows with bias magnitude and optical power, but depends
weakly on scan speed over $1$--$200\,\mathrm{nm/s}$.
A quasi-static surface-trap model links voltage-biased carrier transport
to asymmetric trapping, whose space-charge field shifts the resonance
through the Pockels effect. The model captures this hysteresis across
three cavities through switching between stable steady states.
Space-charge feedback opposes further carrier separation, limiting the
hysteresis gap. The model informs photorefractive mitigation and resonance
control in integrated electro-optic devices.
\end{abstract}

\keywords{thin-film lithium niobate; photorefractive effect; photonic-crystal cavities; electro-optic modulation; surface traps; hysteresis}

\setboolean{displaycopyright}{true}

\section{Introduction}
\label{sec:intro}

Thin-film lithium niobate (TFLN) has emerged as an
increasingly mature platform for integrated electro-optics: its
large Pockels coefficient enables fast, low-voltage modulation, its
wide transparency window supports large optical bandwidth, and
wafer-scale nanofabrication makes dense integration
practical~\cite{Wang2018,Boes2023,Zhu2021}.
These properties have enabled the demonstration of modulators
with bandwidths reaching $100\,\mathrm{GHz}$~\cite{Wang2018},
broadband frequency combs~\cite{Zhang2019,Englebert2026}, and microwave-to-optical
transducers for quantum interconnects~\cite{McKenna2020,Holzgrafe2020}. {}
Advances in lithium niobate nanofabrication~\cite{Zhu2021} have
extended the TFLN platform to include nanostructured resonators such
as photonic-crystal cavities (PhCCs). PhCCs, by combining
wavelength-scale mode volumes with high quality factors, strongly
enhance intracavity optical fields, enabling nonlinear frequency
conversion~\cite{Jiang2018}, compact electro-optic
modulators~\cite{Li2020,Larocque2024}, and photonic
interconnects~\cite{BuzziRouter2026}.

However, the same intracavity field enhancement that
enables these applications also makes PhCCs particularly
susceptible to the photorefractive (PR)
effect~\cite{Ashkin1966}.
The PR effect arises when absorbed photons excite free
carriers that drift under internal and applied electric fields and
fill trap sites asymmetrically, building a space-charge electric
field~\cite{Ashkin1966,Kukhtarev1979}; in bulk crystals these traps
are dopant defects, while in nanophotonic TFLN devices surface and
dielectric-interface traps can play an important
role~\cite{Xu2021,Yin2026}.
The space-charge field shifts the optical resonance via the
Pockels effect. Feedback between the intracavity intensity and trap
occupation can support multiple stable steady states at the same laser
wavelength. These states can be probed by measuring optical transmission
while sweeping the laser wavelength across the resonance. As the wavelength
changes, the cavity follows a continuous sequence of stable states, called
a branch, until that branch loses stability and the cavity switches to
another. Opposite sweep directions can follow different branches and switch
at different wavelengths, producing hysteresis even in the quasi-static
limit.
History-dependent resonance shifts can compromise resonance-voltage
calibration~\cite{Savchenkov2006}, disrupt
wavelength-locking loops~\cite{Leidinger2016}, and introduce
uncertainty into quantum protocols that assume a fixed cavity
coupling strength.

Photorefractive effects on cavity resonances have been
reported across many TFLN resonator geometries.
The effect has been characterized as a function of
wavelength and temperature in whispering-gallery
resonators~\cite{Leidinger2016,Xu2021cryo}, and shown to produce
full-linewidth resonance shifts at intracavity photon numbers of
only a few in photonic-crystal cavities~\cite{Liang2017,Li2019}.
While the PR effect is fundamentally electrical in origin,
its response under an externally applied electric field has not been previously
systematically studied in integrated TFLN resonators.
Without an applied bias, the bulk photovoltaic effect can provide an
internal field that biases carrier migration~\cite{Glass1974}. An applied
voltage changes how photogenerated carriers fill the traps, altering the
photorefractive shift of the cavity resonance.

\begin{figure}[t]
\centering
\includegraphics[width=\linewidth]{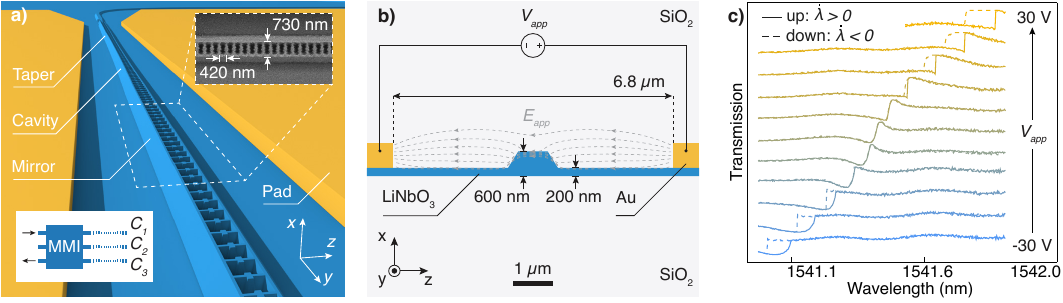}
\caption{Device geometry and voltage-dependent optical
  response. (a) Perspective view of the TFLN photonic-crystal cavity and
  lateral electrodes; the upper inset shows a micrograph of the photonic
  crystal, and the lower inset shows the full $3\times3$ MMI device with tunable
  cavities $C_1$, $C_2$, and $C_3$. Each cavity has a separately
  addressable electrode for voltage tuning; the electrodes are omitted
  from the inset for clarity.
  (b) Cross-section of the oxide-clad rib waveguide and transverse
  applied field. (c) Normalized, vertically offset raw spectra of cavity $C_1$
  at fixed input power and scan speed. Solid and dashed traces denote
  increasing- and decreasing-wavelength sweeps, respectively. Their
  overall displacement shows electro-optic tuning, while the separation
  within each pair shows photorefractive hysteresis.}
\label{fig:device}
\end{figure}

\begin{figure}[t]
\centering
\includegraphics[width=\linewidth]{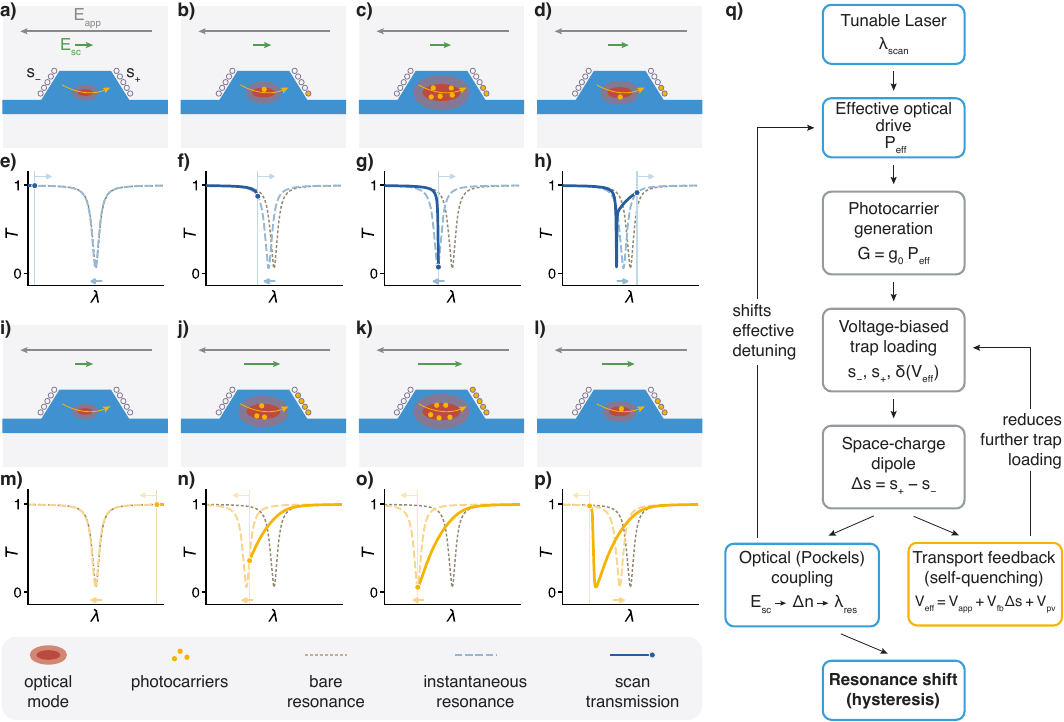}
\caption{Surface-trap photorefractive mechanism underlying
  sweep-direction hysteresis and the corresponding model architecture.
  (a--d) Cross-sectional cavity schematics and (e--h)
  corresponding transmission plots at four stages of an
  increasing-wavelength sweep. The schematics show the optical mode,
  photocarriers, surface traps, and applied and space-charge fields; the
  plots distinguish the bare resonance, instantaneous resonance, and scan
  transmission up to each stage. Trap loading pulls the cavity resonance
  toward the scanning laser, compressing the trace. (i--l) Schematics
  and (m--p) transmission plots for a decreasing-wavelength sweep, in
  which the resonance is dragged away from the scanning laser,
  stretching the trace. (q) STPR model flow
  linking effective optical drive, photocarrier generation, voltage-biased
  trap loading, space charge, and the resonance shift. Blue outlines indicate
  the optical response, gray outlines the carrier dynamics, and gold
  outlines the transport feedback; block colors do not denote scan direction.}
\label{fig:architecture}
\end{figure}

Increasing the bias magnitude widens the gap between resonance
wavelengths measured during increasing- and decreasing-wavelength laser
scans. We interpret this hysteresis with a surface-trap photorefractive
(STPR) model. The cavities are part of a device designed for three-port
pairwise optical routing~\cite{BuzziRouter2026}.
The space-charge field shifts the cavity resonance through
the Pockels effect and feeds back on the drift of photogenerated carriers.
The model reproduces measurements from three cavities fitted independently.
We also test shared-parameter fits across the three cavities and, separately,
across laser scan speeds of $1$--$200\,\mathrm{nm\,s^{-1}}$ for $C_1$.
These measurements do not determine the absolute trapping rates.

\section{Device and Experimental Observations}
\label{sec:device}

We use the three photonic-crystal cavities of a $3\times3$
interferometric switch to study how applied voltage changes
photorefractive hysteresis. We apply DC biases so that the cavity resonances are sufficiently separated in wavelength for each cavity to be characterized independently. Figure~\ref{fig:device} shows the device and
illustrates how we observe hysteresis by comparing transmission spectra
from opposite laser scan directions.

\subsection{Device Structure}

The switch was fabricated on an x-cut TFLN platform
($600\,\mathrm{nm}$ LiNbO$_3$ on
$4.7\,\mu\mathrm{m}$ buried SiO$_2$ on silicon, clad with silicon
dioxide). Its three arms terminate in photonic-crystal nanobeam cavities
$C_1$, $C_2$, and $C_3$. The switch provides independent optical access to
each cavity's resonance.
Light enters through a tapered waveguide into a cavity
section where the beam width is gradually modulated to form a
localized resonant mode, then reflects back off a photonic-crystal
mirror [Fig.~\ref{fig:device}(a)].
More information on the full device can be found
in~\cite{BuzziRouter2026}.
Lateral gold electrodes with a spacing of $6.8\,\mu\mathrm{m}$
[Fig.~\ref{fig:device}(b)] apply a DC bias, held fixed during each laser scan. The bias on the
swept cavity ranges from $-30$ to $+30\,\mathrm{V}$.
At the central wavelength $\lambda_0\approx1541.59\,\mathrm{nm}$,
a linear fit to the low-power voltage sweep gives an effective
quasi-static tuning magnitude
$|\mathrm{d}\nu/\mathrm{d}V|=1.92\pm0.06\,\mathrm{GHz\,V^{-1}}$
($|\mathrm{d}\lambda/\mathrm{d}V|\approx(15.2\pm0.5)\,\mathrm{pm\,V^{-1}}$)~\cite{BuzziRouter2026}.
To determine each cavity's loaded quality factor, we jointly fit its
three output spectra at the two lowest input powers. The fits share a
resonance wavelength and linewidth while allowing different asymmetric
lineshapes caused by interference with the other arms. This gives loaded
quality factors $Q\approx5.4\times10^4$, $5.0\times10^4$, and
$5.1\times10^4$ for $C_1$--$C_3$,
respectively, corresponding to full widths at half maximum of
approximately $28$, $31$, and $30\,\mathrm{pm}$.

\subsection{Measurement Protocol}

We swept a tunable continuous-wave laser across the cavity
resonances in both wavelength directions and recorded a full transmission
spectrum for each sweep. We define the signed scan rate as
$\dot\lambda\equiv\mathrm{d}\lambda/\mathrm{d}t$, positive for increasing
wavelength and negative for decreasing wavelength. The scan speed
$|\dot\lambda|$ ranged from $1$ to $200\,\mathrm{nm\,s^{-1}}$.
After subtracting the common spectral background, we use local Fano
fits to help identify the cavity feature and track its wavelength shift
across neighboring voltage settings. We take the wavelength where the
measured transmission rises most steeply as the resonance position.
The extraction procedure, exclusions, and validation are detailed in Supplementary
Note~\ref{app:tracking}.

For each of the three cavities, we recorded transmission spectra at
laser scan speeds of $2$, $5$, and $20\,\mathrm{nm\,s^{-1}}$, using 16 voltage
settings from $-30$ to $+30\,\mathrm{V}$ in $4\,\mathrm{V}$ increments.
We refer to these measurements as the three-cavity dataset.
We repeated these measurements at five source powers of $-14$, $-9$,
$-4$, $+1$, and $+6\,\mathrm{dBm}$, corresponding to estimated cavity-arm
input powers of $-26.8$, $-21.8$, $-16.8$, $-11.8$, and $-6.8\,\mathrm{dBm}$.
The conversion assumes an $8\,\mathrm{dB}$ grating-coupler loss and the
equal three-way split of the $3\times3$ multimode interferometer (MMI),
$10\log_{10}3\approx4.8\,\mathrm{dB}$.
Each voltage step was followed by a $0.5\,\mathrm{s}$ settling
time before the paired wavelength scans. Further acquisition timing is
given in Supplementary Note~\ref{supp:methods}.
We applied DC bias to cavities $C_2$ and $C_3$ to detune their resonances and spectrally
isolate the response of cavity $C_1$. The separately acquired scan-speed
dataset contains measurements of $C_1$ at $1$, $10$, $50$, and
$200\,\mathrm{nm\,s^{-1}}$, using the same five input powers and
11 voltage settings from $-30$ to $+30\,\mathrm{V}$ in $6\,\mathrm{V}$ increments.
We use the three-cavity dataset for the independent cavity fits.
For $C_1$, we also combine both datasets in a joint fit across all seven
scan speeds. Separately, we test whether the physical parameters determined
from the three-cavity dataset describe the scan-speed dataset. We retain those physical parameters and use the two lowest input powers
to determine a common electro-optic tuning coefficient $\rV$ for all four
additional scan speeds and a separate reference wavelength $\lambda_0$
for each acquisition. Here $\lambda_0$ is the cavity resonance wavelength
at zero applied voltage in the absence of photorefraction. We then compare
the model with the three higher powers without further fitting.
Figure~\ref{fig:device}(c) shows representative raw
transmission spectra across an 11-step voltage sweep, directly
illustrating the resonance shift and sweep-direction hysteresis
before any extraction or fitting is applied.

\subsection{Key Experimental Observations}
\label{sec:obs}

Figure~\ref{fig:device}(c) shows the voltage-enhanced photorefractive
hysteresis for cavity $C_1$ at a representative estimated cavity-arm input power
($-16.8\,\mathrm{dBm}$).
The resonance wavelengths extracted from increasing-wavelength sweeps
($\lambda_\mathrm{res}^\uparrow$) and decreasing-wavelength sweeps
($\lambda_\mathrm{res}^\downarrow$) differ at fixed $\Vapp$, producing a hysteresis gap
$\Delta\lambda = \lambda_\mathrm{res}^\uparrow - \lambda_\mathrm{res}^\downarrow$
that reaches tens of picometers.
The magnitude of the gap generally increases with $|\Vapp|$,
approximately linearly above about $10\,\mathrm{V}$, without evident
saturation up to $30\,\mathrm{V}$.
The hysteresis gap is small near zero bias and increases substantially
under applied voltage.
The gap magnitude also generally increases with optical input power. The
measured range does not show a clear plateau and therefore does not establish
a saturation power.

\section{Surface-Trap Photorefractive Model}
\label{sec:model}

To explain why applied voltage enhances the measured hysteresis,
we model how the electric field biases carrier trapping and how the
trapped charge shifts the cavity resonance. We then use the steady states
of the trap equations to describe why opposite laser scan directions can
produce different resonance wavelengths.

\subsection{Physical Picture}

We develop a surface-trap reduction of the Kukhtarev--Vinetskii
band-transport model~\cite{Kukhtarev1979} tailored to the TFLN PhCC
geometry.
Photogenerated carriers drift in the net transverse electric field
and diffuse across the waveguide. Surface traps at the etched sidewalls
capture the carriers, and unequal trapping on the two sides builds a
space-charge dipole.
Figure~\ref{fig:architecture}(a--p) shows how the space-charge
dipole shifts the cavity resonance differently in the two scan directions.
Panel (q) shows how these mechanisms relate in the STPR model.
Figure~\ref{fig:architecture}(a,e) shows the up-sweep before
the laser reaches resonance, with negligible carrier generation and
little deviation from the bare resonance. As the laser approaches
resonance (b,f), rising intracavity power generates carriers that
load the sidewall traps and build a space-charge dipole, which pulls
the resonance toward the scanning laser and compresses the observed
lineshape (c,g). Once the laser has swept past resonance (d,h), the
intracavity power drops, carrier generation falls off, and the
resonance recoils back toward its unperturbed position. Figure
\ref{fig:architecture}(i--p) shows the down-sweep undergoing the
same sequence in reverse (i,m before resonance; j,n approaching; k,o
at resonance; l,p after), except the dipole now pushes the resonance
away from the laser instead of toward it, stretching rather than
compressing the lineshape. For the illustrated positive-bias case, the net effect
of this screening is that the observed resonance sits at shorter
wavelength than the bare resonance.
The physical picture is formalized in the model of
Fig.~\ref{fig:architecture}(q): the effective optical drive $\Pcav$ sets
the photocarrier generation rate $G=\gzero\Pcav$, the resulting
carrier partition is described by the asymmetry $\delta=p_+-p_-$, where
$p_+$ and $p_-$ are the fractions of generated carriers reaching the two
sidewalls. The trap occupancies $\fp,\fm$ follow Shockley--Read--Hall
(SRH) kinetics. Their difference $\Df=\fp-\fm$ determines the modeled
space-charge field, which in turn affects the optical resonance and
carrier drift. For the fitted scans, we solve these
rate equations at steady state, self-consistently with the cavity response.
In this steady-state description, opposite scan directions can reach
different equilibrium states at the same laser wavelength.

\subsection{Cavity Response}

The optical power inside the cavity drives carrier generation, so
we first calculate how that power varies with laser detuning. The photon lifetime
$\tau_\mathrm{ph} = Q/\omega_0 \approx 40\,\mathrm{ps}$ is more than ten
orders of magnitude shorter than the approximately $1\,\mathrm{s}$
required to scan from the edge of the wavelength window to the cavity
resonance at $2\,\mathrm{nm\,s^{-1}}$, so the
intracavity power follows the instantaneous detuning quasi-statically
(adiabatic approximation). We express this dependence through the effective
optical drive $\Pcav$:
\begin{equation}
  \Pcav = P_\mathrm{in}\left[1 + \left(\Delta_\mathrm{tot}/\gamma\right)^{\!2}\right]^{-1},
  \label{eq:cavity}
\end{equation}
We express detuning and linewidth in wavelength units. Over the narrow
scan window, the frequency-domain Lorentzian is well approximated by this
form, with wavelength half-width $\gamma=\lambda_0/(2Q)$. The total detuning
\begin{equation}
  \Delta_\mathrm{tot} = \lambda_\mathrm{scan} - \lambda_0 - \rV\Vapp + \rV\Vscmax\Df
  \label{eq:detuning}
\end{equation}
combines the laser's detuning from the bare resonance
$\lambda_0+\rV\Vapp$, including the electro-optic (Pockels) shift
from the applied bias, and the screening term
$\rV\Vscmax\Df$ from the field associated with the trap occupancy
imbalance $\Df$ via the resonance-shift coefficient $\Vscmax$. The bare resonance,
$\lambda_0+\rV\Vapp$, includes the electro-optic shift from the applied
bias but excludes the photorefractive shift from trapped charge
($\Vscmax=0$). Its zero-bias value is $\lambda_0$.
Here $P_\mathrm{in}$ is the estimated optical power delivered to the cavity arm.
Equation~\eqref{eq:cavity} uses a unit-peak Lorentzian, so $\Pcav$ is an effective
optical drive rather than the absolute circulating power; the fixed resonant
build-up factor is absorbed into the fitted generation coefficient $\gzero$.

The half-linewidth $\gamma=\lambda_0/(2Q)$ is fixed by the loaded
quality factor extracted from low-power spectra. We determine $Q$ by fitting
the three output spectra simultaneously with Fano lineshapes, which account
for interference between the cavity response and light from the other arms.
For each low-power laser sweep, the three output spectra are fitted with
a common resonance wavelength and linewidth, while allowing separate
background and lineshape coefficients for each output. For each cavity,
we use the median quality factor extracted from the low-power scans at
$2$ and $5\,\mathrm{nm\,s^{-1}}$ to set the model linewidth.

These fits use the measured transmission spectra after the output MMI.
The Lorentzian in Eq.~\eqref{eq:cavity} describes the effective optical
drive; the output transmission is calculated using the full $3\times3$
interferometer model.

\subsection{Carrier Partition: Langevin Function from
  First-Passage Theory}
\label{sec:langevin}

Following the photoexcited-carrier migration and trapping picture of
Xu et al.~\cite{Xu2021}, we treat the transported excitation as an effective
mobile charged species without assigning a specific electronic or ionic
identity. The mobile species perform a biased random walk in the transverse
direction. We define $\Veff$ as the signed carrier-drift voltage, positive for
drift toward the right ($+$) sidewall. The mapping between the applied field
and $\Veff$ depends on the charge sign and microscopic transport mechanism and
is absorbed into this signed definition.
The probability $u(\xi)$ that a single carrier generated at normalized
transverse position $\xi\in[0,1]$ becomes trapped at the right
($+, \xi=1$) sidewall rather than the left ($-, \xi=0$) sidewall is the
first-passage probability satisfying the backward Kolmogorov
equation~\cite{Gardiner2004}:
\begin{equation}
  \frac{d^2u}{d\xi^2} + \Peclet\,\frac{du}{d\xi} = 0,
  \quad u(0)=0,\; u(1)=1,
  \quad \Peclet \equiv \frac{\Veff}{\VT},
  \label{eq:bk}
\end{equation}
Here $\Peclet$ is the carrier P\'eclet number, the dimensionless
ratio of drift to diffusive transport across the waveguide. For
$|\Peclet|\ll1$, diffusion produces nearly symmetric sidewall trapping;
for $|\Peclet|\gg1$, drift dominates and the sign of $\Peclet$ selects
the favored sidewall. We parameterize this ratio by the effective
transport-voltage scale $\VT$, such that $\Peclet=\Veff/\VT$. For conventional
nondegenerate transport, the Einstein relation gives $D/\mu=k_BT/|q|$,
where $D$ is the diffusion coefficient, $\mu$ the carrier mobility,
$k_B$ Boltzmann's constant, $T$ the temperature, and $q$ the carrier charge.
Here, however, $\VT$ is fitted phenomenologically
and may incorporate non-Einstein hopping or trapping dynamics, field geometry,
and the conversion between applied voltage and the microscopic transverse
field; it is not interpreted as the lattice thermal voltage.
The probability of reaching the right sidewall from position $\xi$ is
$u(\xi) = (1 - e^{-\Peclet\,\xi})/(1 - e^{-\Peclet})$.

We assume that carriers are generated uniformly across the waveguide,
so every starting position contributes equally. The probability $u(\xi)$
gives the chance that a carrier generated at position $\xi$ reaches the
right sidewall. Averaging $u(\xi)$ across the waveguide therefore gives the
fraction $p_+$ of all generated carriers reaching that wall:
\[
 p_+=\int_0^1 u(\xi)\,\mathrm{d}\xi.
\]
In this model, every carrier eventually reaches one of the two sidewalls,
so the fraction reaching the left wall is $p_-=1-p_+$. Their difference,
$\delta=p_+-p_-$, measures the preference for trapping at one sidewall over
the other. Substituting the expression for $u(\xi)$ gives
\begin{equation}
  \delta \equiv p_+ - p_-
    = \coth\!\left(\frac{\Veff}{2\VT}\right) - \frac{2\VT}{\Veff}
    \equiv L\!\left(\frac{\Veff}{2\VT}\right),
  \label{eq:langevin}
\end{equation}
where $L(\xi) = \coth\xi - 1/\xi$ is the Langevin function.
Fitting $C_1$ alone, using all five powers and the three original scan
speeds, gives $\VT=0.011\,\mathrm{V}$
(Supplementary Table~\ref{tab:perchannel}). The partition is controlled by
the effective voltage, including transport feedback, rather than by the
applied voltage alone. The sensitivity to centered optical-mode-like
generation profiles is reported in Supplementary
Table~\ref{tab:carrier_profiles}.

\subsection{Trap Kinetics}
\label{sec:traps}

The right ($+$) and left ($-$) sidewall trap occupancy fractions
$\fp(t), \fm(t) \in [0,1]$ evolve by Shockley--Read--Hall
kinetics~\cite{Shockley1952}:
\begin{align}
  \dot{\fp} &= \gzero\Pcav\;\tfrac{1}{2}(1+\delta)\;(1-\fp)
               - \krel\,\fp,
  \label{eq:ode_plus}\\[3pt]
  \dot{\fm} &= \gzero\Pcav\;\tfrac{1}{2}(1-\delta)\;(1-\fm)
               - \krel\,\fm,
  \label{eq:ode_minus}
\end{align}
where the effective photogeneration rate $G=\gzero\Pcav$ is
proportional to the effective optical drive, with coefficient
$\gzero$ in $\mathrm{mW}^{-1}\,\mathrm{s}^{-1}$. The vacant-trap factor
$(1-s_\pm)$ limits further capture as the traps fill. We assume that the
trap-release rate $\krel$ follows an Arrhenius law
$\krel = k_0\,e^{-E_a/k_BT}$, where $k_0$ is the release-rate
prefactor, $E_a$ is the trap-release activation energy, $k_B$ is the
Boltzmann constant, and $T$ is the absolute temperature.
The trap occupancy imbalance $\Df = \fp - \fm$ determines the
space-charge field in the model.
The weak measured dependence on scan speed motivates a quasi-static
approximation. Away from switching points, we assume that the trap
populations relax to local steady state faster than the laser detuning
changes appreciably. Setting
$\dot s_\pm=0$ in Eqs.~\eqref{eq:ode_plus} and~\eqref{eq:ode_minus}
gives
\begin{equation}
 s_\pm^\mathrm{ss}=\frac{a p_\pm}{1+a p_\pm},\qquad
 a=\Rtrap\Pcav,\qquad \Rtrap=\frac{\gzero}{\krel}.
 \label{eq:steadyoccupancy}
\end{equation}
Here $\Rtrap$ has units of $\mathrm{mW^{-1}}$, and $a=G/\krel$ is the
dimensionless ratio of the effective photogeneration rate to the trap-release
rate. The loading-to-release ratio for each sidewall is $a p_\pm$. Each
occupancy grows linearly for $a p_\pm\ll1$ and approaches one for
$a p_\pm\gg1$.
The scale $1/\Rtrap$ is the effective optical drive at which the generation
rate $G$ equals the release rate $\krel$. Each sidewall reaches half
occupancy when $\Rtrap\Pcav p_\pm=1$.
We assume that the two sidewalls can hold equal amounts of trapped charge
when fully occupied. The charge imbalance is then proportional to
$\fp-\fm$. As both sides approach full occupancy, this imbalance and its
contribution to the resonance shift approach zero. Filling the traps
therefore does not necessarily make the hysteresis gap approach a constant value.

\subsection{Voltage Enhancement and Space-Charge Feedback}
\label{sec:couplings}

The applied external voltage is included in the model through
an effective drift voltage
\begin{equation}
  \Veff = \Vapp + \Vpv + \Vfb\,\Df,
  \label{eq:veff}
\end{equation}
where $\Vpv$ is the built-in photovoltaic offset and $\Vfb$ is the
transport-feedback coefficient. The fitted built-in photovoltaic offset
may contain a contribution from the bulk photovoltaic effect, but the
resonance scans alone do not isolate its microscopic origin.
As $|\Veff|$ increases, drift becomes stronger relative to diffusion,
and the two sidewalls receive increasingly unequal fractions of the
photogenerated carriers.

The space-charge field has two effects: it shifts the cavity resonance
through the Pockels effect and modifies the drift of subsequently generated
carriers.
The first effect is a resonance shift through the Pockels effect.
We define the screening term $\epss$ by
\begin{equation}
  \epss = \rV\,\Vscmax\,\Df,
  \label{eq:screening}
\end{equation}
The corresponding PR-induced resonance-wavelength shift relative to the
electro-optic position is $\Delta\lambda_\mathrm{PR}=-\epss$.
Here $\Vscmax \propto \int r_{33}(x)\,E_\mathrm{sc}(x)\,
|\mathcal{E}(x)|^2\,dx$ encodes the spatial overlap of the dominant Pockels
tensor element $r_{33}(x)$, the space-charge field profile $E_\mathrm{sc}(x)$ ($\Df=1$), and the optical mode intensity $|\mathcal{E}(x)|^2$; since $|\Df|\le1$, $\Vscmax$ is the resonance-shift
voltage at maximum trap imbalance, $|\Df|=1$.
The second effect is feedback on carrier transport. The
$\Vfb\Df$ term in Eq.~\eqref{eq:veff} modifies the drift of
subsequently generated carriers.
We restrict $\Vfb\le0$ to represent opposing space-charge feedback.
For $\Vfb<0$, increasing $\Df$ reduces $\Veff$ and the carrier-partition
asymmetry, thereby suppressing further asymmetric trap loading.
\subsection{Quasi-Static Reduction and Branch Selection}
\label{sec:quasistatic}

To predict the resonance along a laser scan, we solve for the
steady-state trap imbalance and effective optical drive together. Subtracting the
two steady-state occupancies gives
\begin{equation}
 \Df=\frac{a\,\delta(\Df)}
 {1+a+\tfrac14a^2[1-\delta(\Df)^2]},\qquad
 a=\Rtrap\Pcav(\lambda_\mathrm{scan},\Df),
 \label{eq:steadyclosure}
\end{equation}
with $\delta(\Df)=L[(\Vapp+\Vpv+\Vfb\Df)/(2\VT)]$ and the
effective optical drive given by Eq.~\eqref{eq:cavity}. At fixed
effective optical drive, negative transport feedback gives a unique stable
steady state (Supplementary Note~\ref{app:fixedpoint}). In the cavity,
however, trapped charge shifts the resonance and thereby changes the
optical drive. This feedback can support two stable steady-state solutions at the
same laser wavelength. As the laser wavelength changes, the cavity remains near its current
equilibrium as long as small disturbances to the trap populations decay.
If that equilibrium becomes unstable, the trap populations evolve toward
another stable state. We use the trap-rate equations,
Eqs.~\eqref{eq:ode_plus} and~\eqref{eq:ode_minus}, to determine which state
is reached. Figure~\ref{fig:steady_hysteresis} illustrates the resulting
hysteresis and how it develops with increasing input power: the scan paths
coincide in panel (a), whereas panels (b,c) show hysteresis.
In the steady-state model, the two rates enter only through their
ratio $\Rtrap=\gzero/\krel$. The fits therefore cannot determine
$\gzero$ and $\krel$ separately.

\section{Numerical Implementation and Fitting}
\label{sec:numerics}

For each measured voltage and input power, we solve
Eq.~\eqref{eq:steadyclosure} over the experimental wavelength window.
The model can have two stable steady-state solutions at a given wavelength.
We initialize each modeled scan on the stable branch connected to the
far-from-resonance state, assuming that the trap populations have relaxed
before the scan begins. As the wavelength changes,
we follow the connected sequence of stable solutions, called a branch.

\begin{figure}[t]
\centering
\includegraphics[width=\linewidth]{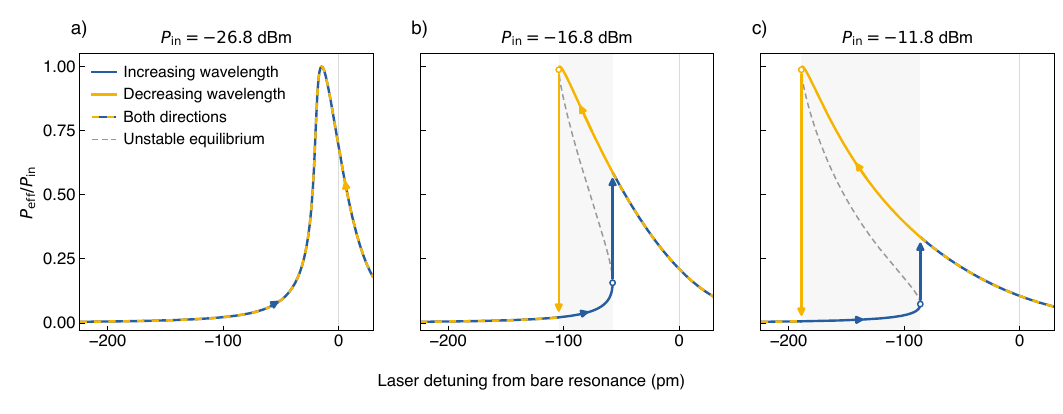}
\caption{Quasi-static hysteresis from stable-branch selection. Normalized
effective optical drive $\Pcav/P_\mathrm{in}$ versus laser detuning from the bare
resonance for the independent $C_3$ fit (Supplementary
Table~\ref{tab:perchannel}) at $\Vapp=+30\,\mathrm{V}$ and estimated
cavity-arm input powers (a)~$-26.8$, (b)~$-16.8$, and
(c)~$-11.8\,\mathrm{dBm}$, using identical axis ranges. Solid blue and yellow
curves follow stable states for increasing and decreasing wavelength,
respectively. Where the paths coincide, blue and yellow alternate along
one continuous curve. Arrows indicate scan directions and jumps between
stable states. Gray dashed curves show unstable equilibria according to
the full two-occupancy Jacobian including cavity feedback. Shaded intervals
contain two stable equilibria, and open circles mark the switching points.
The vertical gray line marks zero detuning from the bare resonance.
The scan paths coincide in (a), while (b,c) exhibit hysteresis. In (c), the two
switches occur at detunings $-86.1$ and $-189.1\,\mathrm{pm}$, respectively.
The plot illustrates the fitted model, not an additional measurement;
no scan-speed magnitude enters the quasi-static calculation.}
\label{fig:steady_hysteresis}
\end{figure}

A steady state is stable if a small change in the trap populations
is followed by a return to their equilibrium values. We call the wavelength
at which the followed branch loses stability a switching point. For example,
during the decreasing-wavelength scan shown in yellow in
Fig.~\ref{fig:steady_hysteresis}(b,c), the system follows the upper stable
branch toward the peak in optical power. Beyond this turning point, at
shorter laser wavelengths, only the lower stable branch remains available,
so the system switches abruptly to it. We use the trap-rate equations,
Eqs.~\eqref{eq:ode_plus} and~\eqref{eq:ode_minus}, to calculate how the
populations evolve during this jump, holding the wavelength just beyond
the switching point. Once they reach the new equilibrium, we continue to model the
scan. Opposite scan directions can therefore follow
different branches and switch at different wavelengths. The stability
test and numerical checks are detailed in Supplementary
Note~\ref{app:integrator}.

We calculate the detector transmission from these solutions using
the three-arm interferometer model described in Supplementary
Note~\ref{supp:waterfall}. Switching between stable states produces a sharp
feature in the detected transmission. Its rising edge provides an
experimentally accessible marker of the transition and lies close to the
calculated switching wavelength for the conditions in
Fig.~\ref{fig:steady_hysteresis}(b,c). We extract this marker from both
measured and simulated transmission using the same procedure
(Supplementary Note~\ref{app:tracking}), and use its separation between
opposite scan directions to quantify the hysteresis gap.

For each acquisition, we fit the device's optical response to the
lowest-power transmission spectra. The eight fitted coefficients describe
the cavity response and the relative amplitudes and phases of light returning
through the three interferometer arms (Supplementary
Note~\ref{supp:waterfall}). We then hold these coefficients fixed while
fitting the resonance positions.

We fit each cavity independently using all five input powers, all
retained voltage settings, and all three scan speeds in the three-cavity dataset. For each
cavity, a single set of physical parameters describes all powers and scan
speeds. The bare electro-optic tuning coefficient $\rV$ is also common to
these measurements, while the reference wavelength $\lambda_0$ is fitted
separately for each scan-speed run to accommodate resonance shifts between
acquisitions, likely caused by temperature drift~\cite{Wang2016ThermoOptic}.

Even at the two lowest input powers, trapped charges shift the cavity
resonance. A linear fit to these scans would therefore mix the bare
electro-optic tuning with photorefractive shifts. We determine $\rV$ and
$\lambda_0$ together with the physical parameters using the full dataset
(Supplementary Note~\ref{supp:methods}).

\begin{table}[t]
\centering
\caption{Parameters and inputs for the independent cavity fits. All quantities
marked ``fitted'' are optimized together using the measured resonance positions.
Each cavity has one set of parameters shared across powers and scan speeds,
except for $\lambda_0$, which is fitted separately for each scan-speed dataset.
The loaded quality factor is measured and the cavity-arm input powers are
estimated; both are held fixed. The interferometer coefficients are fitted
separately to low-power transmission spectra and held fixed during the
resonance-position fits.}
\label{tab:model_parameters}
\small
\begin{tabular}{@{}llll@{}}
\toprule
Symbol & Meaning & Units & Treatment in the fit \\
\midrule
$\Rtrap=\gzero/\krel$ & Trap loading-to-release ratio & $\mathrm{mW^{-1}}$ & Fitted (per cavity) \\
$\VT$ & Carrier-transport voltage scale & V & Fitted (per cavity) \\
$\Vscmax$ & Resonance-shift coefficient & V & Fitted (per cavity) \\
$\Vfb$ & Transport-feedback coefficient & V & Fitted (per cavity) \\
$\Vpv$ & Effective photovoltaic offset & V & Fitted (per cavity) \\
$\rV$ & Bare electro-optic tuning coefficient & $\mathrm{pm\,V^{-1}}$ & Fitted (per cavity) \\
$\lambda_0$ & Zero-bias reference wavelength & nm & Fitted (per scan-speed dataset) \\
$Q$ & Loaded quality factor & --- & Fixed \\
$P_\mathrm{in}$ & Estimated cavity-arm input power & mW & Fixed \\
\bottomrule
\end{tabular}
\end{table}

We fit the model by minimizing a cost function based on the mean squared
differences between measured and calculated resonance wavelengths for both
laser scan directions.
The fitted parameters and fixed inputs are listed in
Table~\ref{tab:model_parameters}. We calculate the hysteresis gaps from the
fitted wavelengths and compare them with the measurements. We use the same
cost function for the independent fits, shared-parameter fits, and model
comparisons. Supplementary Note~\ref{app:integrator} defines the cost function.

To check numerical reliability, we test the sensitivity of the fits to
starting values and numerical resolution (Supplementary Note~\ref{app:integrator}).
To assess parameter robustness, we repeat the fits on resampled datasets
(Supplementary Note~\ref{app:conditional_sensitivity}).

\section{Results}
\label{sec:results}
\subsection{Independent and Joint Fits}

\begin{figure}[t]
\centering
\includegraphics[width=\linewidth]{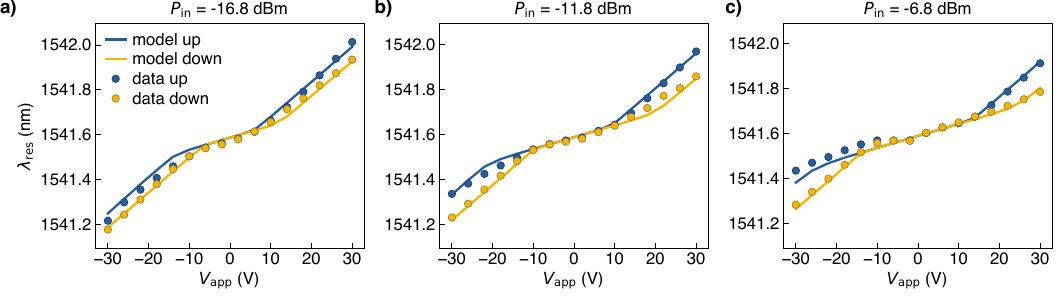}
\caption{Measured and modeled resonance tuning of cavity $C_1$.
  Resonance wavelength across the stepped applied-voltage sweep at
  $2\,\mathrm{nm\,s^{-1}}$ for estimated cavity-arm input powers of
  (a) $-16.8\,\mathrm{dBm}$, (b) $-11.8\,\mathrm{dBm}$, and
  (c) $-6.8\,\mathrm{dBm}$. Blue and gold symbols mark the measured
  cavity-resonance positions during increasing- and decreasing-wavelength
  sweeps, respectively; lines show positions extracted from modeled
  transmission using the same rising-edge procedure
  (RMS residuals $18.7$, $15.7$, and $18.7\,\mathrm{pm}$). The fit includes all five
  powers and all three scan speeds with the numerical scan protocol of
  Sec.~\ref{sec:numerics}.}
\label{fig:K1}
\end{figure}

Figure~\ref{fig:K1}(a--c) shows the measured resonance wavelength of
cavity $C_1$ as the applied voltage is stepped from $-30$ to
$+30\,\mathrm{V}$. At each voltage, separate increasing- and
decreasing-wavelength sweeps at $2\,\mathrm{nm\,s^{-1}}$ provide the two
resonance positions for each of three estimated cavity-arm input powers. At every
power, the resonance shifts with applied voltage, while a direction-dependent
separation is observed between the up- and down-sweep traces. This
separation increases monotonically with estimated cavity-arm input power across the three
measurements.

The STPR model successfully reproduces the overall voltage dependence
and the increase in sweep-direction separation with power, but underestimates
the continued growth of the hysteresis gap at the highest measured power.
Across the three cavities, the root-mean-square (RMS) difference between
modeled and measured resonance wavelengths is $13.2$--$17.7\,\mathrm{pm}$
at $2\,\mathrm{nm\,s^{-1}}$ and $13.9$--$17.3\,\mathrm{pm}$ over all three
scan speeds, evaluated over the three higher input powers
($-16.8$ to $-6.8\,\mathrm{dBm}$).

Independent fits to the three cavities give $\Rtrap=22.5$--$39.0\,\mathrm{mW^{-1}}$,
$\VT=0.011$--$0.170\,\mathrm{V}$, and
$\Vfb/\Vscmax=-1.50$ to $-1.45$. Supplementary Table~\ref{tab:perchannel}
reports the complete parameter sets.

To quantify the effect of sharing parameters, we fit the three
cavities jointly with common $\Rtrap$ and $\VT$, while allowing
$\Vscmax$, $\Vfb$, $\Vpv$, $\rV$, and $\lambda_0$ to remain cavity-specific. Sharing those
two parameters reduces the physical parameter count from 15 to 11, or
from 27 to 23 including $\rV$ and the wavelength offsets. The separately
fitted interferometer coefficients are fixed in both comparisons
(Table~\ref{tab:model_parameters}). Both use the wavelength cost function
defined in Supplementary Note~\ref{app:integrator}, with equal weight
for each cavity and input power.

The joint fit gives an upper-power position RMS of
$15.42\,\mathrm{pm}$ over 283 positions at
$2\,\mathrm{nm\,s^{-1}}$ and $15.73\,\mathrm{pm}$ over all three speeds.
Sharing $\Rtrap$ and $\VT$ increases the cost function value by
$1.4\%$ relative to fitting them independently for each cavity. A common
pair therefore describes these measurements nearly as well as separate
values. Supplementary Table~\ref{tab:universal} and
Fig.~\ref{fig:universal_fit} give the parameters and curves. Supplementary
Fig.~\ref{fig:K3} shows the independent fit and measured branches for cavity $C_3$.

\subsection{Full-Spectrum Comparison}
\label{sec:fullspectra}

\begin{figure}[t]
\centering
\includegraphics[width=\linewidth]{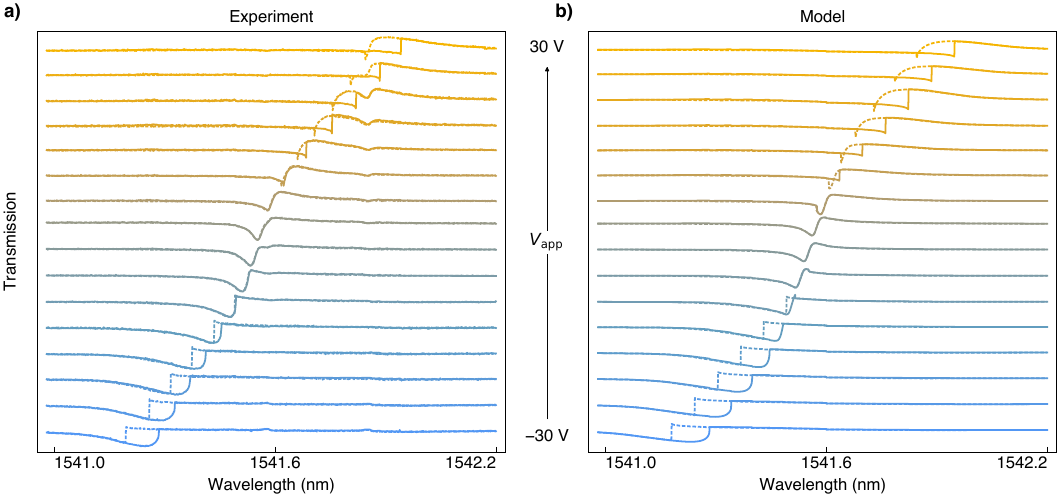}
\caption{Measured and modeled spectra of cavity $C_3$ at
  $2\,\mathrm{nm\,s^{-1}}$ and a $+1\,\mathrm{dBm}$ source setting
  (estimated cavity-arm input power $-11.8\,\mathrm{dBm}$).
  (a) Experiment and (b) steady-state STPR model with an effective
  $3\times3$ readout. Color progresses from blue to gold as the applied
  voltage increases from $-30$ to $+30\,\mathrm{V}$ in $4\,\mathrm{V}$ steps;
  solid and dashed curves denote increasing- and decreasing-wavelength
  sweeps. The four traces at each voltage share one vertical normalization
  and offset. Background processing and optical calibration are detailed
  in Supplementary Note~\ref{supp:waterfall}.}
\label{fig:C3spectra}
\end{figure}

Figure~\ref{fig:C3spectra} compares the measured (a) and modeled (b)
transmission spectra of $C_3$, testing whether the parameters fitted to
resonance positions also reproduce the spectral shapes. We calculate the
transmission using the steady-state cavity response and a three-arm
interferometer model that accounts for interference between the cavity and
two reference arms. The interferometer coefficients are fixed across all
voltages and both scan directions.

The calculated spectra reproduce the main changes in line shape
with applied voltage, with an overall normalized RMS spectral error of
$10.3\%$. The interferometer model and error calculation are described in
Supplementary Note~\ref{supp:waterfall}.

\subsection{Scan-Speed Dependence}
\label{sec:a46validation}

We repeated the power- and voltage-dependent measurements on cavity
$C_1$ at four additional scan speeds ($1$, $10$, $50$, and
$200\,\mathrm{nm\,s^{-1}}$). We jointly fit the $C_1$ measurements from the three-cavity and scan-speed
datasets, covering all seven scan speeds and five powers, using a common
set of physical parameters.
For each sweep direction, the model follows a stable branch and switches
when that branch loses stability; scan speed does not enter the calculation.
The shared parameters are $\Rtrap$, $\VT$, $\Vscmax$, $\Vfb$, and $\Vpv$.
A single bare electro-optic tuning coefficient $\rV$ is shared across all seven speeds; only
the resonance offset $\lambda_0$ is fitted separately for each measurement
run. Supplementary Fig.~\ref{fig:a46residuals} compares the fit with
measurements at the four additional speeds; the acquisition details and
fit parameters are given in Supplementary Note~\ref{app:seven_speed_fit}.

Across all four additional scan speeds, the measured hysteresis gap
retains a similar voltage dependence and does not collapse at the fastest
scan. At $+24\,\mathrm{V}$ and an estimated cavity-arm input power of
$-11.8\,\mathrm{dBm}$, the measured gap is $81$, $81$, $86$, and
$88\,\mathrm{pm}$ at $1$, $10$, $50$, and $200\,\mathrm{nm\,s^{-1}}$,
respectively. The shared fit follows the measured branches and their
voltage-dependent separation, with a pooled position RMS of
$16.7\,\mathrm{pm}$ over the 234 retained positions at the three upper
powers and four additional speeds.

Keeping the physical parameters from the three-cavity dataset fixed
gives a pooled RMS position error of $18.9\,\mathrm{pm}$ for the higher-power
scans at the additional speeds, compared with $16.7\,\mathrm{pm}$ obtained
from the joint fit. This supports the applicability of the quasi-static
model across the measured scan-speed range of
$1$--$200\,\mathrm{nm\,s^{-1}}$, with the same physical parameters describing
the hysteretic response of $C_1$ throughout this range. The test procedure
is detailed in Supplementary Note~\ref{app:joint7}.

\section{Discussion and Conclusion}
\label{sec:discussion}
\label{sec:conclusion}

In the STPR model, applied bias changes the distribution of
photoexcited carriers between the sidewall traps. Unequal accumulation of
these carriers at the two sidewalls creates a space-charge field that shifts
the cavity resonance through the electro-optic effect. At fixed optical
power, negative transport feedback gives the trap populations a unique,
locally stable steady state (Supplementary Note~\ref{app:fixedpoint}). In the
cavity, trap loading changes the detuning and therefore the optical power
driving further loading. This feedback can support two stable steady states
at the same laser wavelength.

The laser scan direction determines which stable branch the system
follows. Increasing- and decreasing-wavelength scans can therefore switch
between states at different wavelengths, producing hysteresis in the measured
resonance position. The modest change in this wavelength separation over a
200-fold range in laser scan speed supports the quasi-static description
used here.

Surface and dielectric-interface traps provide a physical basis for
the voltage-dependent carrier redistribution in the model. Previous
measurements in TFLN microrings showed a strong influence of dielectric
cladding on photorefraction, including suppression after cladding removal
and re-annealing~\cite{Xu2021}. A closely related air-clad TFLN PhCC IQ
modulator exhibits approximately linear voltage tuning without the pronounced
voltage-enhanced hysteresis observed here, despite slower PR-related
transmission drift~\cite{Larocque2024}. Together with the suppression of
surface-state-driven carrier noise by post-fabrication annealing~\cite{Yin2026},
these observations support a contribution from surface and interface
traps. In the STPR model, the applied field redistributes photoexcited
carriers between the traps through Eqs.~\eqref{eq:veff}
and~\eqref{eq:langevin}. At zero applied bias, the fitted photovoltaic
offset can still bias the carrier partition. The offset is an effective fit
parameter, so its value alone cannot identify a microscopic photovoltaic
field.

The fitted ratio $\Vfb/\Vscmax$ compares the transport-feedback
coefficient $\Vfb$ with the resonance-shift coefficient $\Vscmax$. The two coefficients weight the
space-charge field differently, according to optical mode intensity for
$\Vscmax$ and carrier density for $\Vfb$. Their ratio is
therefore not fixed by the model. The fitted coefficients vary across cavities,
with $\Vfb/\Vscmax$ ranging from $-1.50$ to $-1.45$ in the independent fits.
Compensation between these coefficients limits their microscopic
interpretation. Negative $\Vfb$ suppresses further asymmetric trap loading
as charge accumulates. The extent to which this feedback opposes the applied
voltage depends on $\Vfb$ and the trap imbalance.

Trap filling and transport feedback set the power and voltage scales
of the modeled hysteresis. At large $|\Veff|/\VT$, the partition
asymmetry approaches unity as $|\delta|\simeq1-2\VT/|\Veff|$.
The resulting trap imbalance also depends on
power and feedback. The characteristic power
$1/\Rtrap\approx26$--$45\,\mu\mathrm{W}$ describes the effective
intracavity drive at which loading competes with release. The measurements
have not established a saturation power. At sufficiently high optical power,
the model predicts that traps at both sidewalls would approach full
occupancy, reducing their occupancy difference and the resulting
photorefractive resonance shift.

The modeled gap flattens at high voltage, whereas the measured gap
continues to grow approximately linearly with $|\Vapp|$ up to
$30\,\mathrm{V}$ (Fig.~\ref{fig:K1}). The model consequently underestimates
the gap at $|\Vapp|\gtrsim20\,\mathrm{V}$, with larger discrepancies at
higher optical power. Improving agreement through physically motivated
extensions of the model is promising future work. {}

We assessed the variability of the fitted parameters and predicted
resonance positions by repeating the fits on 512 artificial datasets with
deviations from the model comparable in size to those observed experimentally. The predicted resonance
positions changed by a median RMS of $3.0$--$3.5\,\mathrm{pm}$ across the four
fits, while some fitted parameters varied more substantially. The
procedure and resulting parameter ranges are described in Supplementary
Note~\ref{app:conditional_sensitivity}.

We also tested sensitivity to ambiguous resonance assignments and
finite wavelength sampling. Excluding ambiguous positions changed the
predicted curves by at most $1.6\,\mathrm{pm}$ RMS. The sampling test gave
changes of about $1\,\mathrm{pm}$; the procedure and its numerical
limitations are described in the Supplementary Information
(Supplementary Notes~\ref{app:bootstrap} and~\ref{app:conditional_sensitivity}).

Xu et al.~\cite{Xu2021} reported pronounced scan-speed dependence in
oxide-clad lithium-niobate microrings, with a characteristic response on
the order of $10\,\mathrm{ms}$. Differences in optical confinement, carrier
transport, and interface traps may contribute to the weaker speed dependence
observed here. Our steady-state fits constrain the loading-to-release ratio
$\Rtrap$, but cannot determine the individual rates or identify which
processes set the response time.
Multiplying $\gzero$ and $\krel$ by the same factor leaves $\Rtrap$ and the
steady-state response unchanged. Time-resolved measurements are needed to
determine the absolute trap rates and the duration of switching transients.
They would also test whether the trap populations lag behind their local
steady state during a laser scan.

Reducing the loading-to-release ratio $\Rtrap$ offers a route to
mitigating photorefractive shifts by raising the characteristic loading
power. Passivating the etched sidewalls or modifying the cladding could reduce
carrier trapping. In the Shockley--Read--Hall rate
equations, either decreasing $\gzero$ or increasing $\krel$ reduces
$\Rtrap$, although the steady-state fits cannot distinguish these routes.
UV illumination has also been shown to accelerate the photorefractive
response in TFLN microdisks~\cite{Hou2024}, demonstrating that the overall
response kinetics can be modified optically.

A promising application of our model would be to guide feed-forward voltage compensation by predicting
how photorefraction changes the dependence of the resonance wavelength
on applied voltage.
Because the response is hysteretic, the correction would depend on which
stable branch the cavity occupies. Computing stable operating points and
selecting voltages within a known branch required only microseconds in our
numerical tests (Supplementary Note~\ref{app:control_timing}). Time-resolved
measurements would be needed to determine how rapidly this compensation
could operate.

Voltage pulses could potentially switch the cavity between its
coexisting stable states, allowing information to be stored in its optical
response under continuous illumination. Our steady-state fits cannot
determine how quickly these states could be written or how long they
would persist after illumination is removed. A write--wait--read experiment
would directly test these possibilities. Long-lived photorefractive and pyroelectric states reported
in TFLN microresonators~\cite{Ren2025} motivate such an investigation.

We observed voltage-enhanced photorefractive hysteresis in three
thin-film lithium niobate photonic-crystal cavities, studied its dependence on
applied voltage, optical power, and laser scan speed, and developed a
quasi-static surface-trap model to explain the measured behavior. The model
captures the observed switching between stable steady states and reproduces
the main trends in voltage- and power-dependent hysteresis. At estimated
cavity-arm input powers of $-16.8$, $-11.8$, and $-6.8\,\mathrm{dBm}$, the
hysteresis gap reaches about $50$--$150\,\mathrm{pm}$ near $|V|=30\,\mathrm{V}$.
The weak dependence on scan speed over $1$--$200\,\mathrm{nm\,s^{-1}}$ supports
the quasi-static description.

The steady-state model could guide resonance calibration by
accounting for scan history. The model also provides a quantitative tool for investigating the reduction
of trap loading towards mitigation of photorefractive shifts.

\section*{Funding}
This work was supported by the DEVCOM Army Research Laboratory under
Cooperative Agreement No.~W911NF-22-2-0127. A.B.\ acknowledges support
from the Air Force Research Laboratory under Award
No.~FA8750-25-2-0500 and from the Defense Advanced Research Projects
Agency (DARPA) through the NanoWatt Platforms for Sensing, Analysis and
Computation (NaPSAC) program as part of the TOPCHIP project under
subcontract No.~SCON-00005835. M.H.\ acknowledges support from The Novo
Nordisk Foundation (REPEAT, grant no.~NNF24OC0096351).

\section*{Acknowledgments}
The authors thank Steven Nagle and the T.~J. Rodgers RLE Laboratory for
their continued support in developing the experimental setup. The authors thank CSEM, in
particular Homa Zarebidaki, Amir H. Ghadimi, and Hamed Sattari, for
fabricating the chips.

Generative AI tools assisted with portions of code development, data analysis,
mathematical analysis, and manuscript review. The authors reviewed and
validated the AI-assisted work. They retain full responsibility for the
scientific content, results, and conclusions.

\section*{Disclosures}
The authors declare no conflicts of interest.

\section*{Code and Data Availability}
\begingroup
\urlstyle{same}
The experimental spectra and extracted resonance positions are on
Zenodo (\url{https://doi.org/10.5281/zenodo.24206999}).
The resonance-tracking and model-fitting code, fitted parameters, and
conditional robustness analyses are available on GitHub
(\url{https://github.com/sandrobuzzi/Voltage-Enhanced-Photorefraction}).
\endgroup

\noindent See Supplement 1 for supporting content.

\bibliographystyle{unsrtnat}
\bibliography{references}

\clearpage
\onecolumn
\setcounter{section}{0}
\setcounter{secnumdepth}{2}
\titleformat{\subsection}{\normalsize\itshape}{\thesubsection}{0.6em}{}
\renewcommand{\theHsection}{supp.\arabic{section}}
\renewcommand{\theHsubsection}{supp.\arabic{section}.\arabic{subsection}}
\setcounter{table}{0}
\renewcommand{\thetable}{S\arabic{table}}
\renewcommand{\tablename}{Supplementary Table}
\renewcommand{\theHtable}{supp.\arabic{table}}
\setcounter{figure}{0}
\renewcommand{\thefigure}{S\arabic{figure}}
\renewcommand{\figurename}{Supplementary Figure}
\renewcommand{\theHfigure}{supp.\arabic{figure}}
\setcounter{equation}{0}
\renewcommand{\theequation}{S\arabic{equation}}
\renewcommand{\theHequation}{supp.\arabic{equation}}

\newcommand{\suppnote}[2]{%
  \refstepcounter{section}%
  \section*{Supplementary Note \arabic{section}: #1}%
  \label{#2}%
}

\begin{center}
  {\LARGE\bfseries Supplementary Information}\\[5mm]
  {\large\bfseries Voltage-Enhanced Photorefractive Hysteresis in
   Thin-Film Lithium Niobate Photonic-Crystal Cavities}\\[5mm]
  Alessandro Buzzi$^{1,*}$, Bjarke A. Frederiksen$^{1,2}$,
  Mikkel Heuck$^{2}$, Dirk R. Englund$^{1}$, and
  Dashiell L. P. Vitullo$^{1,3}$\\[3mm]
  $^{1}$Research Laboratory of Electronics, Massachusetts Institute of
  Technology, Cambridge, Massachusetts 02139, USA\\
  $^{2}$Department of Electrical and Photonics Engineering, Technical
  University of Denmark, Kgs. Lyngby 2800, Denmark\\
  $^{3}$DEVCOM Army Research Laboratory, Adelphi, Maryland 20783, USA\\[2mm]
  $^{*}$Corresponding author: abuzzi@mit.edu
\end{center}

\vspace{6mm}
\noindent\textbf{Contents}\\[1mm]
\hspace*{4mm}Supplementary Note 1: Fitting procedure and numerical methods\\
\hspace*{4mm}Supplementary Note 2: Fit results and model comparison\\
\hspace*{4mm}Supplementary Note 3: Extended-sweep validation\\
\hspace*{4mm}Supplementary Note 4: Resonance tracking\\
\hspace*{4mm}Supplementary Note 5: Fit robustness and identifiability\\
\hspace*{4mm}Supplementary Note 6: Carrier-partition models and self-quenching\\
\hspace*{4mm}Supplementary Note 7: Measured spectra and model waterfalls
\vspace{6mm}

\suppnote{Fitting Procedure and Numerical Methods}{supp:methods}

We describe the measurement sequence, simulated scan protocol,
and fitting procedure used to relate the measured resonance positions
to the STPR parameters.

\subsection{Acquisition and Simulated Scan Sequence}

For the $2$, $5$, and $20\,\mathrm{nm\,s^{-1}}$ measurements, the
laser scans $1540$--$1543\,\mathrm{nm}$ in the decreasing-wavelength
direction before the increasing-wavelength direction at each bias setting. The
voltage program runs from $-30$ to $+30\,\mathrm{V}$ and back in
$4\,\mathrm{V}$ steps, while the saved archive contains one pair of scans
at each of the 16 distinct voltages. The independent $1$, $10$, $50$,
and $200\,\mathrm{nm\,s^{-1}}$ measurements scan
$1541$--$1542\,\mathrm{nm}$ in the increasing-wavelength direction
before the decreasing-wavelength direction at each bias setting. Both acquisition
routines return the laser to $1550\,\mathrm{nm}$ between individual
scans; sweep-completion waits and laser positioning leave a seconds-long
interval between the paired sweeps. The quasi-static calculation uses the
recorded wavelength windows and sweep directions. We assume that, before
each scan begins, the trap populations have relaxed to their stable
equilibrium far from resonance. Returning the laser to $1550\,\mathrm{nm}$ places it far from the cavity
resonances, reducing cavity-enhanced excitation during the seconds-long
interval between scans and allowing trapped carriers to relax. The model does not infer a dark
relaxation time or simulate optical excitation during the laser return.

\subsection{Steady-State Solver and Fitting Cost Function}
\label{app:integrator}

For each input power, we calculate the mean squared difference
between measured and modeled resonance wavelengths over all retained
voltage settings, scan speeds, and scan directions. We then average these
errors with equal weight for each power,
\begin{equation}
 \mathcal{J}(\bm\theta)=\frac{1}{N_P}\sum_{p=1}^{N_P}
 \frac{1}{N_{\lambda,p}}\sum_{i=1}^{N_{\lambda,p}}
 \left[\lambda_{\mathrm{sim},pi}(\bm\theta)
       -\lambda_{\mathrm{obs},pi}\right]^2,
 \label{eq:balancedcost}
\end{equation}
Here $\bm\theta$ denotes the set of parameters being fitted, and
$\lambda_{\mathrm{sim},pi}$ and $\lambda_{\mathrm{obs},pi}$ are the modeled
and measured resonance wavelengths for observation $i$ at input-power
setting $p$. $N_P$ counts the input powers, and $N_{\lambda,p}$ counts the
retained observations at each power. Wavelength differences are
expressed in picometers. All retained points within a power have equal
weight; powers with more retained points therefore do not dominate the
fit. The independent cavity and seven-speed fits use all five powers.
The shared-cavity fit averages this cost function equally across the three
cavities. Model comparisons use the same cost function and retained data.
The hysteresis gap measures the separation between the sharp transmission
transitions in the two laser scan directions. We determine it from the
extracted resonance wavelengths after fitting; it does not enter the cost
function separately.

To test the model against the scan-speed dataset, we hold
$\Rtrap$, $\VT$, $\Vscmax$, $\Vfb$, and $\Vpv$ fixed at their values fitted to
$C_1$ in the three-cavity dataset. Using only the two lowest input powers of
the scan-speed dataset, we fit one common electro-optic tuning coefficient
$\rV$ and a separate reference wavelength $\lambda_0$ for each acquisition.
The three-arm interferometer coefficients remain fixed. We evaluate the
three higher powers without further fitting.

We solve the steady-state equation, Eq.~\eqref{eq:steadyclosure}, numerically.
We determine stability from the Jacobian of the trap-rate equations,
Eqs.~\eqref{eq:ode_plus} and~\eqref{eq:ode_minus}. This $2\times2$ matrix
contains the derivatives of the two trapping rates with respect to the two
trap populations, including the dependence of cavity power on trapped
charge. A steady state is stable when both eigenvalues have negative real
parts, so small disturbances decay. We locate switching points
where the followed steady state loses stability. For each of the three independent cavity fits and the seven-speed $C_1$
fit, we run the optimization from different starting sets of parameter
values. We compare the resulting fits and retain the solution with the
smallest cost function value. Fourteen starting sets per fit are
optimized, followed by six additional local starts, refinement on a finer
grid, and a derivative-free local search. Every accepted solution must
produce all required resonance wavelengths along complete stable scan
paths. These checks reduce dependence on the starting values but do not
guarantee that we have found the global minimum.

For the fit tables, we also report the mean squared wavelength error over
the specified subset of measurements
\begin{equation}
  \mathrm{MSE}_{\mathrm{subset}}(\bm\theta)
  = \frac{1}{N}
    \sum_{\mathrm{pw},\,\mathrm{dir},\,i}
    \bigl[
      \lambda_\mathrm{sim}(V_i;\bm\theta)
     -\lambda_\mathrm{obs}(V_i)
    \bigr]^2 \!\times 10^6
  \quad[\mathrm{pm}^2],
  \label{eq:cost}
\end{equation}
Here $N$ is the number of included measurements, and the wavelength differences
are expressed in nanometers; the factor $10^6$ converts the squared differences
to $\mathrm{pm}^2$.
The error is evaluated over the points stated in each table (the 16 voltage steps,
the three upper power levels, and two sweep directions for the
$2\,\mathrm{nm\,s^{-1}}$ summaries). Unlike the fitting cost function, this descriptive error pools the stated points without equalizing the powers. The parameters $\Rtrap$ and $\VT$ are log-parameterized to enforce positivity and improve conditioning.

For numerical stability, the Langevin function is evaluated via its
Taylor expansion for small arguments ($|\xi| < 10^{-2}$), avoiding
catastrophic cancellation in $\coth\xi - 1/\xi$.

For every parameter set tested during fitting, we simulate each wavelength
scan by following stable steady states, for which both eigenvalues of the
Jacobian have negative real parts. After a loss of stability, the trap populations must settle into a definite
stable equilibrium before the scan continues. Parameter sets are excluded
if this transition cannot be determined or if any required modeled resonance
position is missing. For the three independent cavity fits and the seven-speed fit, an
independent calculation checks 700 complete scan paths
across all five powers, including the original $2\,\mathrm{nm\,s^{-1}}$
and additional $1$ and $200\,\mathrm{nm\,s^{-1}}$ scans.
Fixed-wavelength integration of the SRH equations verifies 141 jumps,
including every loss of stability to oscillatory disturbances (Hopf
bifurcation) in those paths and representative switching points where stable
and unstable branches meet and terminate (folds). In every checked transition,
the populations relax to the selected stable equilibrium.
The independent calculation and refinement of the 801-point
geometry grid to 1601 points change the selected transmission markers by
less than $0.009\,\mathrm{pm}$, indicating the 801-point grid has sufficient resolution. All 440 additional-acquisition scan starts
have a unique attracting state at the recorded wavelength-window endpoint
for the selected seven-speed fit. Absolute rate scaling supplies only an
arbitrary time unit in the jump checks; it does not determine a physical
switching time.

\subsection{Photorefractive Curvature in the Low-Power Calibration}
\label{app:rvcorrection}

Calibrating the bare electro-optic tuning coefficient $\rV$ and resonance offset
$\lambda_0$ by a linear regression on the
lowest-power series (estimated cavity-arm input powers of $-26.8$ and
$-21.8\,\mathrm{dBm}$) assumes negligible photorefractive curvature.
The steady-state model retains a photorefractive shift at these powers,
and the extracted positions can still differ between sweep directions.

Residual photorefraction can therefore bias the linear calibration.
For this reason $\rV$ and $\lambda_0$ are
fitted jointly with the STPR parameters through the model itself, using all five
powers, in both the independent cavity fits and the shared seven-speed fit. The jointly fitted
$\rV$ is $15.6$--$15.7\,\mathrm{pm\,V^{-1}}$ for the independent $C_1$--$C_3$
fits and $15.36\,\mathrm{pm\,V^{-1}}$ for the shared seven-speed fit.

\subsection{Computational Speed and Prospective Resonance Control}
\label{app:control_timing}

We measured how long the model takes to update the cavity resonance
after a small change in laser wavelength. Using the fitted parameters of all
three cavities at three input powers, each update on a known stable branch
took approximately $1.3$--$1.4\,\mu\mathrm{s}$ on an Apple M4 Pro. We also
tested the reverse calculation: finding the applied voltage needed to reach
a chosen cavity resonance while holding the laser wavelength and input power
fixed. For a target on a known stable branch, this took approximately
$18\,\mu\mathrm{s}$. These timings exclude initialization and measure
computation time, not how quickly the physical cavity responds.

\suppnote{Fit Results and Model Comparison}{supp:fitcomparison}

We report the per-cavity and shared-cavity parameter sets, then
compare how changes to the carrier partition and feedback affect the fits.

\subsection{Fit Parameters}

\begin{table}[htbp]
\centering
\caption{Quasi-static STPR parameters fitted to the three-cavity dataset, including
all five input powers and scan speeds of $2$, $5$, and $20\,\mathrm{nm\,s^{-1}}$. The jointly fitted $\rV$ values are 15.56, 15.68, 15.67$\,\mathrm{pm\,V^{-1}}$ for $C_1$--$C_3$, with one resonance offset
per speed. Fixed measured loaded quality factors are $5.4\times10^4$,
$5.0\times10^4$, and $5.1\times10^4$. The mean squared wavelength error (MSE) uses the three upper
powers at $2\,\mathrm{nm\,s^{-1}}$. CV is the descriptive coefficient of
variation across three point estimates, not parameter uncertainty.
Ratios are calculated before rounding.}
\label{tab:perchannel}
\small
\begin{tabular}{@{}lrrrrrrr@{}}
\toprule
Ch. & $\Rtrap$ & $\VT$ & $\Vscmax$ & $\Vfb$ & $\Vpv$ & $\Vfb/\Vscmax$ & MSE \\
 & (mW$^{-1}$) & (V) & (V) & (V) & (V) & {} & (pm$^2$) \\
\midrule
$C_1$ & 39.01 & 0.011 & $+19.18$ & $-28.76$ & $+0.411$ & $-1.499$ & 314.9 \\
$C_2$ & 22.46 & 0.101 & $+20.91$ & $-31.08$ & $+0.257$ & $-1.486$ & 246.9 \\
$C_3$ & 22.88 & 0.170 & $+20.68$ & $-29.93$ & $+0.350$ & $-1.447$ & 175.3 \\
\midrule
CV\,[\%] & 34 & 85 & 5 & 4 & --- & --- & --- \\
\bottomrule
\end{tabular}
\end{table}

\begin{figure}[t]
\centering
\includegraphics[width=\linewidth]{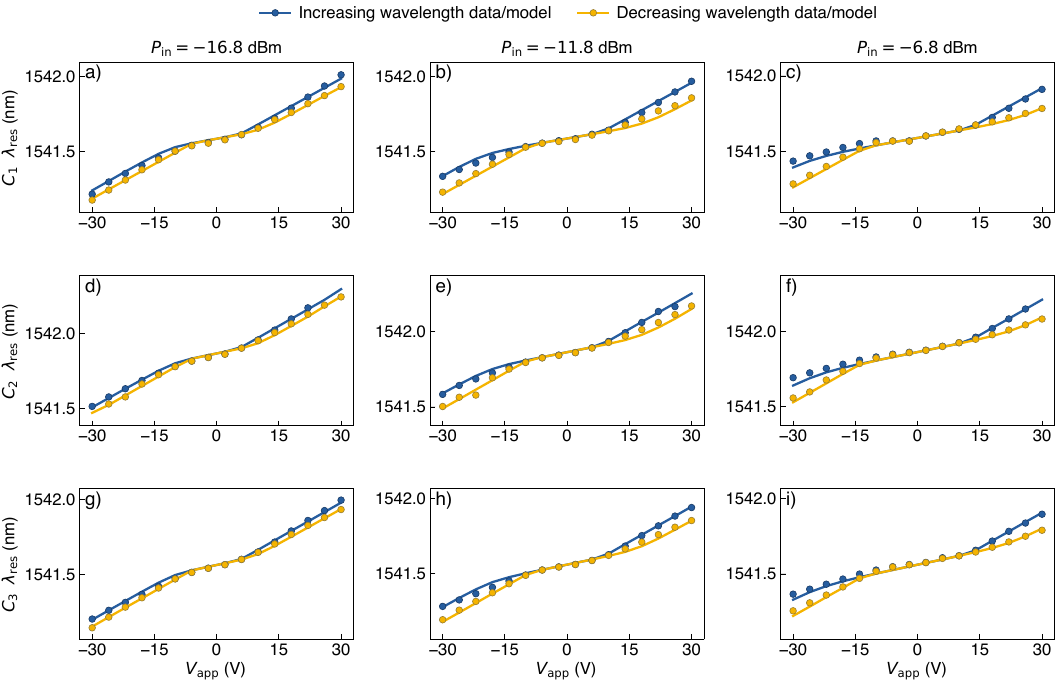}
\caption{Shared quasi-static fit across the three cavities. Measured
  resonance positions (points) and predictions from the shared-parameter
  fit (lines) at $2\,\mathrm{nm\,s^{-1}}$ are shown for cavities
  $C_1$--$C_3$ (rows) and the three highest input powers (columns). Blue and
  gold denote increasing- and decreasing-wavelength sweeps, respectively.
  The fit shares $\Rtrap$ and $\VT$ across all cavities while
  retaining cavity-specific $\Vscmax$, $\Vfb$, and $\Vpv$, and gives an overall
  RMS residual of $15.42\,\mathrm{pm}$ over the 283 displayed points at
  $2\,\mathrm{nm\,s^{-1}}$ (mean squared wavelength error
  $237.7\,\mathrm{pm}^2$).}
\label{fig:universal_fit}
\end{figure}

\begin{table}[htbp]
\centering
\caption{Shared quasi-static fit using the same wavelength cost function as the
independent fits. $\Rtrap$ and $\VT$ are shared; resonance-shift, transport-feedback, and wavelength
calibration parameters are cavity-specific. The fitted $\rV$ values are
15.28, 15.75, 15.67$\,\mathrm{pm\,V^{-1}}$, with one offset per cavity and speed.
The optical readouts and measured $Q$ values are fixed.
The reported mean squared wavelength error includes the three highest
input powers and all three scan speeds ($2$, $5$, and $20\,\mathrm{nm\,s^{-1}}$).}
\label{tab:universal}
\small
\begin{tabular}{@{}llrl@{}}
\toprule
Scope & Parameter & Value & Unit \\
\midrule
Shared & $\Rtrap$ & 24.71 & mW$^{-1}$ \\
Shared & $\VT$ & 0.100 & V \\
\midrule
$C_1$ & $\Vscmax$ & $+23.05$ & V \\
$C_1$ & $\Vfb$ & $-32.29$ & V \\
$C_1$ & $\Vpv$ & $+0.288$ & V \\
$C_2$ & $\Vscmax$ & $+20.53$ & V \\
$C_2$ & $\Vfb$ & $-30.38$ & V \\
$C_2$ & $\Vpv$ & $+0.454$ & V \\
$C_3$ & $\Vscmax$ & $+19.48$ & V \\
$C_3$ & $\Vfb$ & $-29.45$ & V \\
$C_3$ & $\Vpv$ & $+0.422$ & V \\
\midrule
All & Mean squared wavelength error & 247.3 & pm$^2$ \\
\bottomrule
\end{tabular}
\end{table}

Sensitivity of the fitted transport scale to its allowed range and
to sharing parameters across cavities is examined in Supplementary
Note~\ref{app:gapweight}. The fitted photovoltaic offsets are effective
parameters that can also absorb voltage calibration and model mismatch;
they do not by themselves measure intrinsic photovoltaic fields.

\begin{figure}[t]
\centering
\includegraphics[width=\linewidth]{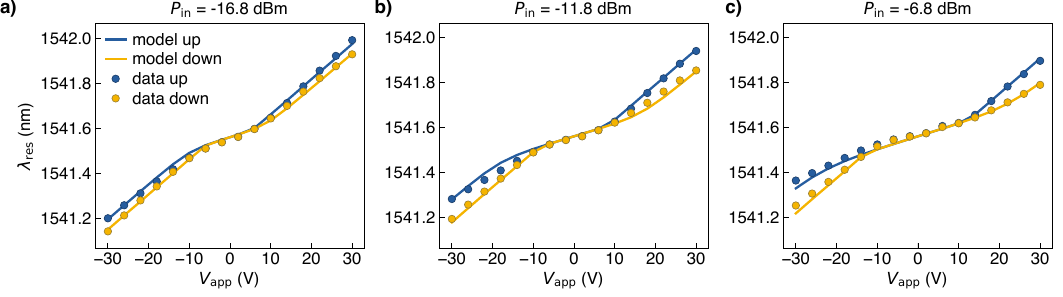}
\caption{Resonance wavelength vs.\ applied voltage for cavity $C_3$.
  Symbols and lines as in Fig.~\ref{fig:K1}.
  Mean squared wavelength error: $175.3\,\mathrm{pm^2}$ ($13.24\,\mathrm{pm}$ RMS).}
\label{fig:K3}
\end{figure}

\subsection{Model Comparison}
\label{sec:comparison}

We test how the assumed distribution of photogenerated carriers and the
feedback of trapped charge on carrier transport affect agreement with the
measurements. We compare the Langevin partition, which assumes uniform
carrier generation across the waveguide, with a partition assuming generation
at its center, $\delta=\tanh(\Veff/4\VT)$. Separately, we remove transport
feedback by setting $\Vfb=0$. Each cavity's loaded $Q$ remains fixed at its
measured value. All variants use the same full optical readout, common tracker, masks,
parameter bounds, and equal-power wavelength cost function. Relative to
Langevin, the point-source cost function changes by $C_1$: $-1.7\%$, $C_2$: $-1.7\%$, $C_3$: $+0.4\%$.
The upper-power position errors are compared in Table~\ref{tab:comparison}.
The point-source partition does not consistently lower the full fitting
cost function; we retain Langevin because it follows the spatially uniform
generation assumption.

Some no-feedback candidates lose stability through a Hopf
bifurcation but do not relax to the proposed destination equilibrium in
the fixed-wavelength integration. We exclude these candidates from an
equilibrium-branch comparison and restrict the no-feedback search to
paths that switch at folds. Within this restricted search, removing
feedback increases the lowest-found all-power wavelength
cost function by $115$--$172\%$.
For $C_1$, the RMS over the two lowest powers changes from
$16.0$ to $29.8\,\mathrm{pm}$
after refitting. This comparison tests whether one parameter set describes
the full power range; it does not measure the direct feedback strength at
low power or establish a unique microscopic mechanism. Previously explored
parameter sets and fresh perturbations are scored under the common cost function,
and eight distinct starting sets per model and cavity are
optimized, followed by grid refinement and three additional local starts. All reported parameter sets must produce complete simulated scans
that follow stable equilibria and yield every required resonance position. These are lowest-found
verified fits, not certified global minima.

\begin{table}[htbp]
\centering
\caption{Matched quasi-static structural comparisons: position MSE
[pm$^2$] at $2\,\mathrm{nm\,s^{-1}}$ on the three upper powers.
All five powers and all three speeds enter the common wavelength cost function.
The Langevin baseline is the independent fit in Table~\ref{tab:perchannel}. The measured $Q$ values, masks, calibration
policy and physical bounds are held fixed except for setting $\Vfb=0$
in the no-feedback model. Reported candidates have verified attracting
continuations; the fold-only restriction and multistart search are described above.
The text reports the corresponding changes in the full fitting cost function.}
\label{tab:comparison}
\small
\begin{tabular}{@{}lrrr@{}}
\toprule
Model & $C_1$ & $C_2$ & $C_3$ \\
\midrule
Langevin [Eq.~\eqref{eq:langevin}] & 314.9 & 246.9 & 175.3 \\
Point-source $\tanh$ & 302.1 & 220.4 & 166.6 \\
No feedback ($\Vfb=0$) & 635.2 & 517.1 & 447.9 \\
\bottomrule
\end{tabular}
\end{table}

\suppnote{Extended-Sweep Validation}{supp:extendedvalidation}

We examine whether one set of physical parameters can describe the
measurements across all seven scan speeds, then fit the slowest scans on
their own. These comparisons use the measured resonance positions to
estimate parameters. A separate transfer test holds the physical parameters
from the three-cavity dataset fixed and tests them against the scan-speed
dataset, as detailed in Supplementary Note~\ref{app:joint7}.

\subsection{Shared Seven-Speed Fit}
\label{app:seven_speed_fit}

At each additional scan speed, the protocol covered the same five input
powers and 11 voltage steps from $-30$ to $+30\,\mathrm{V}$ in
$6\,\mathrm{V}$ increments, while
$C_2$ and $C_3$ were held at fixed biases of $20$ and $40\,\mathrm{V}$,
respectively, to prevent their resonances from crossing $C_1$ during
tuning and thereby improve resonance tracking.
The additional-speed spectra, hysteresis gaps, and fits reported here
all use the scan-speed dataset.
The number of wavelength samples per $1\,\mathrm{nm}$ sweep decreases
with scan speed ($1000$, $1000$, $200$, and $100$ points at $1$, $10$, $50$, and
$200\,\mathrm{nm\,s^{-1}}$, that is, $1$, $1$, $5$, and $10\,\mathrm{pm}$ per
point, compared with the approximately $30\,\mathrm{pm}$ linewidth).
The $200\,\mathrm{nm\,s^{-1}}$ spectra are therefore sampled most coarsely;
manual-label validation does not establish a corresponding loss of tracking
accuracy (Supplementary Note~\ref{app:tracking}).

Table~\ref{tab:a46} reports a single fit to all five optical powers
and all seven scan speeds for cavity $C_1$. The five physical parameters
$\Rtrap$, $\VT$, $\Vscmax$, $\Vfb$, and $\Vpv$ have common values across
the complete dataset, together with a single bare electro-optic tuning coefficient
$\rV$. Only the zero-bias resonance wavelength $\lambda_0$ varies between
measurement runs, allowing for wavelength offsets. The fit has 13 free
parameters: five physical parameters, one $\rV$, and seven offsets. The
steady-state equations contain no scan-speed-dependent rate parameter.
The two acquisitions
retain their separately measured loaded quality factors,
$5.4\times10^4$ for the original $2$, $5$, and $20\,\mathrm{nm\,s^{-1}}$
scans and $5.1\times10^4$ for the additional $1$, $10$, $50$, and
$200\,\mathrm{nm\,s^{-1}}$ scans, and their recorded wavelength windows.
The fit uses the wavelength cost function in Supplementary
Note~\ref{app:integrator}.

The all-power position RMS is $16.32\,\mathrm{pm}$ over 859
positions; the upper-power RMS is $16.81\,\mathrm{pm}$ over 521
positions and the gap RMS is $20.36\,\mathrm{pm}$ over 257 pairs.
For the four additional speeds in Supplementary Fig.~\ref{fig:a46residuals},
the position and gap RMS values are $16.70$ and
$19.23\,\mathrm{pm}$ over 234 positions and
114 pairs. At $200\,\mathrm{nm\,s^{-1}}$ they are
$17.86$ and $16.10\,\mathrm{pm}$, respectively.

\begin{figure}[t]
\centering
\includegraphics[width=\linewidth]{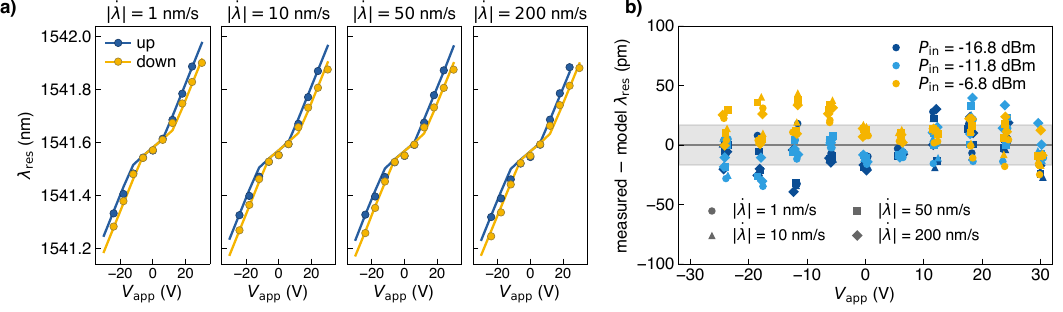}
\caption{Scan-speed dependence and model residuals for cavity
  $C_1$. (a) Measured resonance positions (symbols) and the shared quasi-static STPR
  fit (lines) during increasing- (blue) and decreasing-wavelength (gold)
  sweeps at an estimated cavity-arm input power of
  $-16.8\,\mathrm{dBm}$. The fit uses all seven $C_1$ scan speeds and all
  five powers, including the four speeds shown. (b) Differences between the
  measured and modeled positions (measured minus modeled) for the three upper estimated cavity-arm input
  powers at the four displayed speeds. Color denotes power, marker shape
  denotes speed, and the shaded band marks the $16.7\,\mathrm{pm}$
  pooled RMS residual over $234$ positions. The modeled positions are the
  steepest rising edges extracted from simulated transmission with the same
  procedure used for the measured spectra.}
\label{fig:a46residuals}
\end{figure}

\begin{table}[htbp]
\centering
\caption{Shared quasi-static fit to cavity $C_1$ across
all seven scan speeds and five input powers, as used in
Supplementary Fig.~\ref{fig:a46residuals}. The five physical parameters and $\rV$ (top) are common to
all speeds; only the fitted wavelength offset (bottom) varies by run.
RMS uses the three upper powers, and $N$ gives retained/available
resonance positions at those powers. All rows belong to one joint fit.
The three-cavity and scan-speed datasets retain their separately measured
loaded quality factors.}
\label{tab:a46}
\small
\setlength{\tabcolsep}{6pt}
\begin{tabular}{@{}rrrrrr@{}}
\toprule
$\Rtrap$ & $\VT$ & $\Vscmax$ & $\Vfb$ & $\Vpv$ & $\rV$ \\
(mW$^{-1}$) & (V) & (V) & (V) & (V) & (pm/V) \\
\midrule
36.17 & 0.011 & $+19.24$ & $-29.19$ & $+0.002$ & 15.36 \\
\bottomrule
\end{tabular}
\par\vspace{3mm}
\begin{tabular}{@{}lrrrr@{}}
\toprule
Acquisition & Speed & $\lambda_0$ & RMS & $N$ \\
 & (nm/s) & (nm) & (pm) & \\
\midrule
Original & 2 & 1541.588 & 17.3 & 96/96 \\
 & 5 & 1541.585 & 16.8 & 95/96 \\
 & 20 & 1541.581 & 16.6 & 96/96 \\
\midrule
Additional & 1 & 1541.580 & 15.3 & 58/66 \\
 & 10 & 1541.569 & 17.5 & 59/66 \\
 & 50 & 1541.567 & 16.0 & 58/66 \\
 & 200 & 1541.565 & 17.9 & 59/66 \\
\bottomrule
\end{tabular}
\end{table}

\subsection{Independent Fit at \texorpdfstring{$1\,\mathrm{nm\,s^{-1}}$}{1 nm/s}}

We also fit the $1\,\mathrm{nm\,s^{-1}}$ measurements at all five powers
on their own, to assess how well the model describes the slowest scans
without requiring the same parameters to describe the faster scans.
The physical and calibration parameters are estimated using only these
measurements, with the wavelength cost function described in Supplementary
Note~\ref{app:integrator}. Figure~\ref{fig:a46tuning} shows the resulting
curves. Eight starting sets were optimized and three additional local
starts were checked; the solution with
the smallest cost function value is reported.

\begin{figure}[t]
\centering
\includegraphics[width=\linewidth]{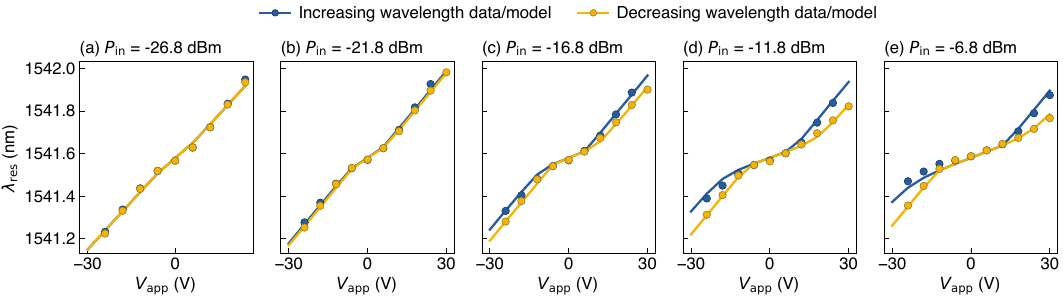}
\caption{Independent quasi-static fit to cavity $C_1$ from the scan-speed
  dataset described in the main text at
  $1\,\mathrm{nm\,s^{-1}}$, using the tracking method described in
  Supplementary Note~\ref{app:tracking_method}, with median agreement of
  $2.0\,\mathrm{pm}$ with the manual reference positions, following the convention of
  Fig.~\ref{fig:K1}. Symbols: measured up-sweep (blue) and down-sweep (gold)
  resonance wavelengths at the five estimated cavity-arm input powers ($-26.8$, $-21.8$,
  $-16.8$, $-11.8$, and $-6.8\,\mathrm{dBm}$, left to right). Lines: STPR model
  on the measured voltage grid. The parameters are
  $\Rtrap=26.28\,\mathrm{mW^{-1}}$, $\VT=0.171\,\mathrm{V}$,
  $\Vscmax=21.65\,\mathrm{V}$, $\Vfb=-29.17\,\mathrm{V}$, and
  $\Vpv=-0.143\,\mathrm{V}$, with $\rV=14.92\,\mathrm{pm\,V^{-1}}$
  and $\lambda_0=1541.5789\,\mathrm{nm}$.
  Mean squared wavelength error: $181.3\,\mathrm{pm^2}$
  ($13.47\,\mathrm{pm}$ RMS over the three upper powers;
  $13.22\,\mathrm{pm}$ over all five).}
\label{fig:a46tuning}
\end{figure}

\suppnote{Resonance Tracking}{app:tracking}

The same transmission-edge extraction is used for the
three-cavity dataset, the scan-speed dataset, and
the simulated spectra. It identifies a reproducible feature of the detected
transmission rather than inferring an optical pole from each trace.

\subsection{Extraction Procedure and Missing Assignments}
\label{app:tracking_method}

For each power, acquisition, and sweep direction, we subtract the
per-wavelength 40th percentile of the transmission across voltage settings.
This removes features that remain approximately stationary as the cavity
resonance moves with applied voltage. We smooth each residual spectrum
with a Gaussian of $3\,\mathrm{pm}$ standard deviation, with a minimum
width of $0.6$ wavelength samples, and locate positive maxima of its
wavelength derivative. A local parabolic interpolation gives the position
between samples. The reported coordinate is this steepest rising edge,
not the center or half-maximum crossing of a fitted line shape.

Candidate edges must exceed three times the robust derivative-noise
scale and have local transmission contrast at least five times the
intensity-noise scale. We estimate both noise scales from fluctuations
within each spectrum, using median-based measures to reduce the influence
of large spectral features. We retain at most eight candidates per spectrum,
separated by at least $15\,\mathrm{pm}$. A generalized Fano fit in a
$\pm100\,\mathrm{pm}$ neighborhood supplies a line-shape quality score;
its fitted center is not used as the resonance coordinate. Across voltage
steps, we select a sequence that favors strong, well-shaped edges and
smooth wavelength motion. The expected voltage trend is estimated from
the spectra themselves, independently for measured and simulated data.
The transition scale is $60\,\mathrm{pm}$ per voltage step, increasing
with the square root of the number of steps when a trace is skipped.
The algorithm may leave a spectrum unassigned; it does not interpolate a
replacement position or use the other sweep direction to supply one.

Two decreasing-wavelength traces in the original $C_1$
acquisition, at $-11.8\,\mathrm{dBm}$, $2\,\mathrm{nm\,s^{-1}}$,
and $+14$ and $+18\,\mathrm{V}$, require reviewed feature selections.
Inspection of the recorded spectra identifies a local resonance edge that
the automatic assignment misses or replaces with a later rising flank.
For these two traces, we identify the feature window manually and apply
the same smoothed-derivative calculation within that window. The windows
are chosen from the spectra, without using model predictions. The original
automatic coordinates and flags are retained in the data record; both fits
that reuse these measurements use the reviewed coordinates.

We exclude seventeen increasing-wavelength spectra clipped by
source-laser mode hops (sixteen for $C_2$ and one for $C_3$). Spectra for
which the tracking procedure returns no resonance position are also omitted
from the fits. The independent fits contain 477, 456, and 477 positions for
$C_1$, $C_2$, and $C_3$, respectively; the seven-speed fit contains 859
positions. The raw spectra, extracted positions, and exclusion masks are
retained in the analysis record.

An assignment is flagged as ambiguous if forbidding it increases
the best complete tracking-path score by less than $0.5$. This flag measures
competition between candidate sequences; it is not a calibrated probability.
Finite flagged positions remain in the fits. We also refit after
excluding them as a separate sensitivity check
(Supplementary Note~\ref{app:bootstrap}).

\subsection{Validation and Model Readout}

We selected 427 spectra across the four additional scan speeds,
prioritizing cases flagged as uncertain by an earlier diagnostic. We used
the manual labels for these spectra to develop and assess the extraction method.
The common tracker
returns a position for all 427 spectra, with two ambiguity flags. Its median
absolute difference from the manual positions is $2.0\,\mathrm{pm}$,
the RMS difference is $4.0\,\mathrm{pm}$, and the largest difference is
$18.6\,\mathrm{pm}$. These labels informed development of the procedure
and are not an independent test set.

The modeled transmission is sampled on the experimental wavelength
and voltage grids and passed through the identical extraction procedure.
Its voltage trend is estimated from its own spectra, without using the
measured positions to select model features. All retained experimental
positions must have finite modeled counterparts. A candidate parameter
set is rejected if the scan path cannot be completed on attracting states
or the tracker cannot assign a required modeled position. A sharp falling
switch can coexist with a broad rising flank: the reported coordinate
remains the selected rising edge in that case. The fitted wavelength offset
can accommodate a common calibration shift, but the marker should not be
interpreted as an intrinsic cavity pole or as every branch-switching point.

\suppnote{Fit Robustness and Identifiability}{supp:robustness}

We assess how perturbing the measured data changes the fitted parameters
and predicted resonance wavelengths. We also test predictions at additional
scan speeds and examine which parameters can be determined from
steady-state data.

\subsection{Search Bounds and Steady-State Identifiability}

The physical bounds are $0.01\le\Rtrap\le100\,\mathrm{mW^{-1}}$,
$0.01\le\VT\le10\,\mathrm{V}$, $0.1\le\Vscmax\le300\,\mathrm{V}$,
$-400\le\Vfb\le0\,\mathrm{V}$, and $-5\le\Vpv\le5\,\mathrm{V}$.
The selected fits lie inside these bounds, but the independent $C_1$
and seven-speed fits give $\VT\simeq0.011\,\mathrm{V}$, close to its lower
bound. Sensitivity to this choice is examined in Supplementary
Note~\ref{app:gapweight}. Common rescaling of $\gzero$ and
$\krel$ is an exact invariance of the steady-state equations: separate
absolute rates cannot be recovered even from noise-free static data.
Practical compensation among $\Rtrap$, $\VT$ and the two voltage coefficients
further limits interpretation of the fitted point estimates.

\subsection{Sensitivity to Ambiguous Assignments}
\label{app:bootstrap}

We repeated the three independent cavity fits and the seven-speed fit after excluding the finite
positions flagged as ambiguous by the common tracker. This removes 21,
7, and 1 positions from the independent $C_1$, $C_2$, and $C_3$ fits,
and 23 from the seven-speed fit. The predicted positions change by
$1.60$, $0.10$, less than $0.01$, and less than $0.01\,\mathrm{pm}$ RMS,
respectively, evaluated on the original retained position sets. This
checks sensitivity to competing feature assignments without treating the
tracking-path score as a statistical confidence level.

\subsection{Conditional Resampling and Tracking Sensitivity}
\label{app:conditional_sensitivity}

\textit{Residual resampling.} We assessed parameter sensitivity by generating 512 artificial datasets for
each of the three independent cavity fits and the shared seven-speed fit.
We first calculated the residual between each measured resonance position
and the selected model prediction. To construct an artificial dataset, we
randomly retained or reversed the signs of those residuals and added them
to the model predictions. Residuals at four adjacent voltage levels within
an acquisition shared one random sign across optical powers and sweep
directions, so nearby errors varied together. This construction adapts
wild residual resampling~\cite{Wu1986} by assigning a common random sign
within each block, as in the wild cluster bootstrap~\cite{Cameron2008}.
The four-voltage block definition is a choice for this sensitivity analysis.

We then refitted the five physical parameters, the common $\rV$, and the
wavelength offsets using the same bounds and equal-power wavelength cost function
as the original fit, with the optical coefficients held fixed. For each
artificial dataset, we refined the original parameter set and two distinct
alternatives selected by their error on that dataset, followed by a separate
least-squares refinement. Additional searches along parameter-coordinate
and random directions were applied to every dataset. A final check of
regularly spaced draws and parameter extremes changed the percentile
interval endpoints by at most $2.1\%$ of their preceding interval widths.
Following the same draws that defined those endpoints gave a conservative
maximum change of $5.7\%$. Lower-error valid solutions were retained;
no artificial dataset was discarded or replaced because of a difficult fit.
These checks assess numerical stability, not uniqueness or global optimality.

\textit{Parameter ranges and predicted positions.} The reported ranges span the 2.5th to 97.5th percentiles of the resulting
parameter estimates. They describe sensitivity to residual perturbations
under fixed input-power calibration, loaded quality factors, optical readout,
extraction settings, reviewed feature selections, exclusion masks, and scan
initialization. They are conditional on the model and resampling scheme,
not confidence intervals for microscopic trap parameters. For $C_1$, $C_2$,
$C_3$, and the seven-speed $C_1$ fit, respectively, the median RMS changes
in predicted positions are $3.17$, $3.50$, $3.05$, and $3.14\,\mathrm{pm}$;
their 97.5th percentiles are $5.36$, $5.69$, $4.58$, and
$4.66\,\mathrm{pm}$. Correlations between $\VT$ and $\Vfb/\Vscmax$ are
$0.675$, $0.869$, $0.960$, and $0.857$, indicating compensation between
the transport scale and the voltage-coefficient ratio.

The transport scale approaches its lower search bound in 77 of 512
$C_1$ refits ($15.0\%$) and 72 seven-speed refits ($14.1\%$). Their
lower percentile limits are therefore constrained by the search range.
No other fitted parameter approaches a bound within $10^{-4}$ of its
transformed search-range width; $R$ and $\VT$ are tested in logarithmic
coordinates. Sensitivity to the allowed transport scale and to sharing it
across cavities is assessed separately in Supplementary Note~\ref{app:gapweight}.

\textit{Wavelength sampling and numerical checks.} A separate 256-refit sampling-resolution study perturbed only the scan-speed
dataset, using correlated common and opposite-direction offsets bounded
by half the recorded wavelength-grid spacing. Relative to the adopted
nominal curve, the median RMS prediction change is $0.970\,\mathrm{pm}$
and its 97.5th percentile is $0.984\,\mathrm{pm}$. Further refinement of
the unperturbed fit changes its curve by $0.962\,\mathrm{pm}$, so the
sampling result is comparable to the numerical refinement scale and does
not establish a precise physical uncertainty. Relative to that refined
unperturbed curve, the sampling-refit changes have a median RMS of
$0.290\,\mathrm{pm}$ and a 97.5th percentile of $1.004\,\mathrm{pm}$.
Sensitivity to ambiguous assignments is assessed in Supplementary
Note~\ref{app:bootstrap}; separate power and quality-factor scenarios follow below.

All 2,304 selected refits return every required transmission marker.
Doubling the geometry grid from 801 to 1601 points changes any extracted
wavelength by at most $0.028\,\mathrm{pm}$. An independent calculation
checks 6,480 scan trajectories from 36 parameter sets spanning the fitted
parameter and prediction extremes. All 1,355 tested fixed-wavelength
integrations relax to the selected equilibrium, and the extracted wavelengths
agree within $0.014\,\mathrm{pm}$. This check does not integrate every jump
for every resample.

Two extreme seven-speed resamples admit a second stable equilibrium at the
recorded beginning of the $1\,\mathrm{nm\,s^{-1}}$, $-21.8\,\mathrm{dBm}$,
$+30\,\mathrm{V}$ decreasing-wavelength scan. Following that alternative
branch gives no extractable rising edge in the recorded window, whereas
the adopted branch reproduces the measured feature. Both resamples are
retained. The analysis therefore remains conditional on the stated
far-from-resonance initialization; it does not assume a unique equilibrium
at every recorded scan boundary. These ranges describe the extracted-position
fit, not uncertainty bands for complete spectra, and do not include
unmodeled spectral physics.

\begin{table}[H]
\centering
\small
\setlength{\tabcolsep}{4pt}
\caption{Conditional parameter ranges from 512 correlated residual refits per dataset, with input power, loaded $Q$ and optical readout fixed. Entries give the selected value followed by the 2.5th--97.5th percentiles. The lower $\VT$ limits for $C_1$ and its seven-speed fit are constrained by the search bound.}
\label{tab:steady_resampling}
\begin{tabular}{lccc}
\toprule
Dataset & $R$ ($\mathrm{mW}^{-1}$) & $V_{\mathrm{tr}}$ (V) & $V_{\mathrm{fb}}/V_{\mathrm{sc}}$ \\
\midrule
$C_1$ & $39.01\,[29.03, 45.20]$ & $0.011\,[0.010, 0.056]$ & $-1.499\,[-1.542, -1.414]$ \\
$C_2$ & $22.46\,[18.06, 25.48]$ & $0.101\,[0.058, 0.371]$ & $-1.486\,[-1.578, -1.260]$ \\
$C_3$ & $22.88\,[19.60, 26.20]$ & $0.170\,[0.086, 0.382]$ & $-1.447\,[-1.541, -1.266]$ \\
$C_1$, seven speeds & $36.17\,[28.17, 41.63]$ & $0.011\,[0.010, 0.139]$ & $-1.517\,[-1.547, -1.381]$ \\
\bottomrule
\end{tabular}
\end{table}

\begin{table}[H]
\centering
\small
\setlength{\tabcolsep}{4pt}
\caption{Additional conditional ranges from the same residual refits as Table~\ref{tab:steady_resampling}. Entries give the selected value followed by the 2.5th--97.5th percentiles; all parameters are in volts.}
\label{tab:steady_resampling_couplings}
\begin{tabular}{lccc}
\toprule
Dataset & $V_{\mathrm{sc}}$ & $V_{\mathrm{fb}}$ & $V_{\mathrm{pv}}$ \\
\midrule
$C_1$ & $19.18\,[18.13, 21.51]$ & $-28.76\,[-30.61, -27.16]$ & $0.411\,[0.088, 0.786]$ \\
$C_2$ & $20.91\,[19.74, 24.38]$ & $-31.08\,[-33.44, -28.68]$ & $0.257\,[-0.025, 0.515]$ \\
$C_3$ & $20.68\,[19.17, 23.04]$ & $-29.93\,[-31.67, -27.23]$ & $0.350\,[0.058, 0.616]$ \\
$C_1$, seven speeds & $19.24\,[18.22, 22.04]$ & $-29.19\,[-30.52, -27.53]$ & $0.002\,[-0.225, 0.220]$ \\
\bottomrule
\end{tabular}
\end{table}

\begin{table}[H]
\centering
\small
\setlength{\tabcolsep}{4pt}
\caption{Conditional calibration ranges from the same residual refits. Entries give the selected value followed by the 2.5th--97.5th percentiles. Each fit shares $r_V$ across its scan speeds; repeated entries denote one fitted coordinate.}
\label{tab:steady_resampling_calibration}
\begin{tabular}{llcc}
\toprule
Dataset & Speed ($\mathrm{nm\,s^{-1}}$) & $r_V$ ($\mathrm{pm\,V^{-1}}$) & $\lambda_0$ (nm) \\
\midrule
$C_1$ & 2 & $15.56\,[15.31, 15.78]$ & $1541.5898\,[1541.5854, 1541.5946]$ \\
$C_1$ & 5 & $15.56\,[15.31, 15.78]$ & $1541.5859\,[1541.5814, 1541.5910]$ \\
$C_1$ & 20 & $15.56\,[15.31, 15.78]$ & $1541.5838\,[1541.5790, 1541.5887]$ \\
\midrule
$C_2$ & 2 & $15.68\,[15.39, 16.01]$ & $1541.8712\,[1541.8650, 1541.8767]$ \\
$C_2$ & 5 & $15.68\,[15.39, 16.01]$ & $1541.8662\,[1541.8616, 1541.8710]$ \\
$C_2$ & 20 & $15.68\,[15.39, 16.01]$ & $1541.8646\,[1541.8613, 1541.8678]$ \\
\midrule
$C_3$ & 2 & $15.67\,[15.46, 15.88]$ & $1541.5639\,[1541.5585, 1541.5690]$ \\
$C_3$ & 5 & $15.67\,[15.46, 15.88]$ & $1541.5605\,[1541.5557, 1541.5653]$ \\
$C_3$ & 20 & $15.67\,[15.46, 15.88]$ & $1541.5562\,[1541.5516, 1541.5603]$ \\
\midrule
$C_1$, seven speeds & 1 & $15.36\,[15.12, 15.52]$ & $1541.5802\,[1541.5763, 1541.5837]$ \\
$C_1$, seven speeds & 2 & $15.36\,[15.12, 15.52]$ & $1541.5884\,[1541.5831, 1541.5931]$ \\
$C_1$, seven speeds & 5 & $15.36\,[15.12, 15.52]$ & $1541.5847\,[1541.5800, 1541.5892]$ \\
$C_1$, seven speeds & 10 & $15.36\,[15.12, 15.52]$ & $1541.5695\,[1541.5651, 1541.5740]$ \\
$C_1$, seven speeds & 20 & $15.36\,[15.12, 15.52]$ & $1541.5810\,[1541.5754, 1541.5862]$ \\
$C_1$, seven speeds & 50 & $15.36\,[15.12, 15.52]$ & $1541.5670\,[1541.5629, 1541.5713]$ \\
$C_1$, seven speeds & 200 & $15.36\,[15.12, 15.52]$ & $1541.5654\,[1541.5602, 1541.5688]$ \\
\bottomrule
\end{tabular}
\end{table}

\textit{Input-power and linewidth sensitivity.} A common multiplicative error in the input-power calibration is
exactly absorbed by an inverse rescaling of $R$: changing every power by a
factor $c_P$ and replacing $R$ by $R/c_P$ leaves the steady-state predictions
unchanged, provided the transformed parameter remains within the search
bounds. Numerical checks at common power shifts of $\pm1\,\mathrm{dB}$ give
identical positions; these shifts are illustrative scenarios, not measured
calibration uncertainties. The inferred generation-to-release ratio therefore depends on the absolute
power calibration even when the fit is unchanged.

We separately refitted all four datasets after common loaded-$Q$
shifts of $\pm5\%$ and $\pm10\%$, keeping the masks, optical readouts,
bounds, and wavelength cost function fixed. The predicted positions change by
$1.4$--$1.7\,\mathrm{pm}$ RMS for the $5\%$ shifts and
$2.2$--$3.3\,\mathrm{pm}$ RMS for the $10\%$ shifts.
Some fitted parameters change much more than the predicted curves. For
example, reducing $Q$ by $10\%$ in the $C_2$ fit changes $\VT$ from
$0.101$ to $0.263\,\mathrm{V}$. Similar resonance curves therefore do not
establish a precisely determined transport scale. All 16 reported refits converge and provide every required
modeled marker on a complete attracting path. When a shifted-$Q$ starting
point lacks a required marker, we screen alternative starting sets for
complete scan paths. We check the nominal parameters and starting sets
with both small and large transport scales, with six additional perturbed
starts for the $10\%$ shifts. These are hypothetical calibration
scenarios, not measured uncertainties in $Q$.

\subsection{Frozen-Parameter Transfer to the Additional Scans}
\label{app:joint7}

A separate transfer test holds the five physical parameters from the
$C_1$ fit to the three-cavity dataset fixed. Using only the two lowest input powers, we fit
one common electro-optic tuning coefficient $\rV$ across the four additional
scan speeds and a separate reference wavelength $\lambda_0$ for each
acquisition. This gives $\rV=15.32\,\mathrm{pm\,V^{-1}}$.
At the three higher powers, which do not enter this calibration, the RMS
errors are $15.8$, $18.8$, $18.7$, and $21.7\,\mathrm{pm}$ at
$1$, $10$, $50$, and $200\,\mathrm{nm\,s^{-1}}$, respectively, with
a pooled RMS of $18.9\,\mathrm{pm}$. The comparison is conditional on
the fixed optical readout calibrated in the primary analysis; it is not
an independent test of the complete extraction and calibration workflow.

\subsection{Sensitivity of the Transport Scale}
\label{app:gapweight}

To assess how precisely the measurements determine the shared
transport scale, we fix $\VT$ at $0.01$, $0.03$, $0.1$, $0.3$, and
$1\,\mathrm{V}$ and refit the remaining parameters. Each calculation uses
the same wavelength cost function and retained observations as the shared-cavity
fit, with three starting parameter sets. Over $0.01$--$0.3\,\mathrm{V}$,
the cost function increases by at most $3.5\%$ and the predicted resonance
wavelengths change by at most $3.2\,\mathrm{pm}$ RMS relative to the
shared-cavity fit. At $1\,\mathrm{V}$, these changes are $13.8\%$ and
$5.8\,\mathrm{pm}$. Thus substantially different transport scales can
describe similar resonance curves, although the larger tested value gives
poorer agreement. These fixed-value tests assess sensitivity; the tested
range is not a confidence interval.

\suppnote{Carrier-Partition Models and Self-Quenching}{app:langevin}

We compare transverse carrier-generation profiles and test how
transport feedback shapes the steady trap response.

\subsection{Uniform, Mode-Like, and Point-Source Carrier Partitions}

For a normalized transverse generation profile $G(\xi)$, with
$\xi\in[0,1]$ and $\int_0^1G(\xi)\,d\xi=1$, the partition asymmetry is
\begin{equation}
  \delta_G = 2\int_0^1G(\xi)u(\xi)\,d\xi-1,
  \label{eq:profile_partition}
\end{equation}
where $u(\xi)=(1-e^{-\Peclet\xi})/(1-e^{-\Peclet})$ is the right-sidewall
first-passage probability. Uniform generation, $G(\xi)=1$, gives the
exact Langevin result $\delta=L(\Veff/2\VT)$ used in the main-text
model. A centered point source, $G(\xi)=\delta_\mathrm{D}(\xi-1/2)$,
instead gives
\begin{equation}
  \delta_\mathrm{point}=\tanh\!\left(\frac{\Veff}{4\VT}\right)
  =\tanh(\chi/2),
  \qquad \chi\equiv\frac{\Veff}{2\VT}.
  \label{eq:point_partition}
\end{equation}

To estimate the sensitivity to optical localization without introducing
an additional fitted profile parameter, we evaluate
Eq.~\eqref{eq:profile_partition} for two representative symmetric,
centered profiles: $G_\mathrm{s}(\xi)=2\sin^2(\pi\xi)$ and a normalized
Gaussian intensity profile with full width at half maximum equal to half the waveguide width.
At the representative drift-dominated values $\chi=10$, $20$, and
$30$, the uniform profile gives
$\delta=0.900$, $0.950$, and $0.967$, the Gaussian gives $0.978$, $0.992$, and
$0.995$, the sine-squared profile gives $0.991$, $0.999$, and $1.000$, and the
centered point source is effectively unity. The mode-like profiles therefore
increase $\delta$ by $2.9$--$10.1\%$ relative to uniform generation, with the
largest increase at the smallest $\chi$, while remaining within $2.2\%$ of the
point-source limit. The assumed transverse profile has a few-percent effect on
the nearly saturated partition at $\chi\gtrsim20$, and the uniform-generation
model provides the conservative value among these centered profiles.

\begin{table}[htbp]
\centering
\caption{Sensitivity of the carrier partition asymmetry $\delta$ to
  the transverse photocarrier-generation profile at representative
  effective drift voltages. Values in parentheses give the increase relative to
  uniform generation.}
\label{tab:carrier_profiles}
\small
\begin{tabular}{@{}lccc@{}}
\toprule
Generation profile & $\chi=10$ & $\chi=20$ & $\chi=30$ \\
\midrule
Uniform (Langevin) & 0.900 & 0.950 & 0.967 \\
Gaussian & 0.978 (8.7\%) & 0.992 (4.4\%) & 0.995 (2.9\%) \\
$2\sin^2(\pi\xi)$ & 0.991 (10.1\%) & 0.999 (5.2\%) & 1.000 (3.4\%) \\
Centered point source & 1.000 (11.1\%) & 1.000 (5.3\%) & 1.000 (3.4\%) \\
\bottomrule
\end{tabular}
\end{table}

\subsection{Uniqueness and Stability at Fixed Optical Drive}
\label{app:fixedpoint}

Hold $\Vapp$ and $\Pcav>0$ fixed, and write $G=\gzero\Pcav>0$,
$k=\krel>0$, $x=\Df$, and
$\delta(x)=L[(\Vapp+\Vpv+\Vfb x)/(2\VT)]$, with $\VT>0$ and $\Vfb<0$.
The steady occupancies are $s_\pm^\mathrm{ss}=Gp_\pm/(Gp_\pm+k)$,
where $p_\pm=(1\pm\delta)/2$. Their difference gives the exact relation
\begin{equation}
  x^\mathrm{ss}=F(x^\mathrm{ss}),\qquad
  F(x)=\frac{4Gk\,\delta(x)}{(G+2k)^2-G^2\delta(x)^2}.
  \label{eq:fixedpoint}
\end{equation}

\textit{Existence and uniqueness.} The right-hand side increases with
$\delta$, whereas $\delta'(x)<0$ because $L'>0$ and $\Vfb<0$.
Thus $F$ is continuous and decreasing, with $|F|<1$ since both steady
occupancies lie in $(0,1)$. Consequently, $x-F(x)$ is strictly increasing,
negative at $x=-1$, and positive at $x=1$, giving exactly one physical
steady state.

\textit{Stability.} Differentiating both trap equations gives the
$2\times2$ Jacobian $J$ at a steady state. Its off-diagonal entries are
nonnegative because $\delta'<0$ and $1-s_\pm^\mathrm{ss}>0$, and its row sums obey
\begin{equation}
  J\begin{pmatrix}1\\1\end{pmatrix}
  =-\begin{pmatrix}Gp_++k\\Gp_-+k\end{pmatrix}.
\end{equation}
Every Gershgorin disc therefore lies strictly in the left half-plane,
so the steady state is locally asymptotically stable. At $G=0$, the unique
steady state is $s_\pm=0$, with decay rate $k$.

This proof assumes externally fixed $\Pcav$. In the coupled cavity model,
$\Pcav$ depends on $\Df$ through detuning, so optical feedback can permit
multiple steady states. The occupancy bounds $0\le s_\pm\le1$ still imply
$|\Df|\le1$, independently of uniqueness.

\subsection{Cavity Feedback and Directional Branches}
\label{app:cavity_branches}

Dividing the numerator and denominator in Eq.~\eqref{eq:fixedpoint}
by $k^2$ gives
Eq.~\eqref{eq:steadyclosure}, with $a=G/k=\Rtrap\Pcav$.
The fixed-power uniqueness proof does not select the branch of the
coupled optical system. For that purpose, we differentiate the original
two occupancy equations at fixed laser wavelength, allowing both
$\Pcav(\Df)$ and $\delta(\Df)$ to vary. A locally attracting steady state
has both Jacobian eigenvalues in the left half-plane. The full Jacobian
equals $\krel$ times a matrix depending only on $\Rtrap$ and the other
steady-state parameters, so positive common rate scaling changes relaxation
times but not stability signs or fixed-wavelength relaxation paths.

The incoming attracting branch is followed until it loses stability.
The next state is selected by relaxation of the SRH rate equations, rather
than by an assumed free-energy minimum or the fixed-power slope alone.
Figure~\ref{fig:steady_hysteresis} shows how hysteresis develops as
the input power increases. Other
tested conditions can lose stability differently;
the numerical rule checks the full Jacobian rather than assuming every
jump occurs at a fold. The switching interval in Fig.~\ref{fig:steady_hysteresis}
is distinct from
the up/down resonance-position gap, which uses the selected rising edges
of the detected transmission features.

\suppnote{Measured Spectra and Model Waterfalls}{supp:waterfall}

\begin{figure}[t]
\centering
\includegraphics[width=\linewidth]{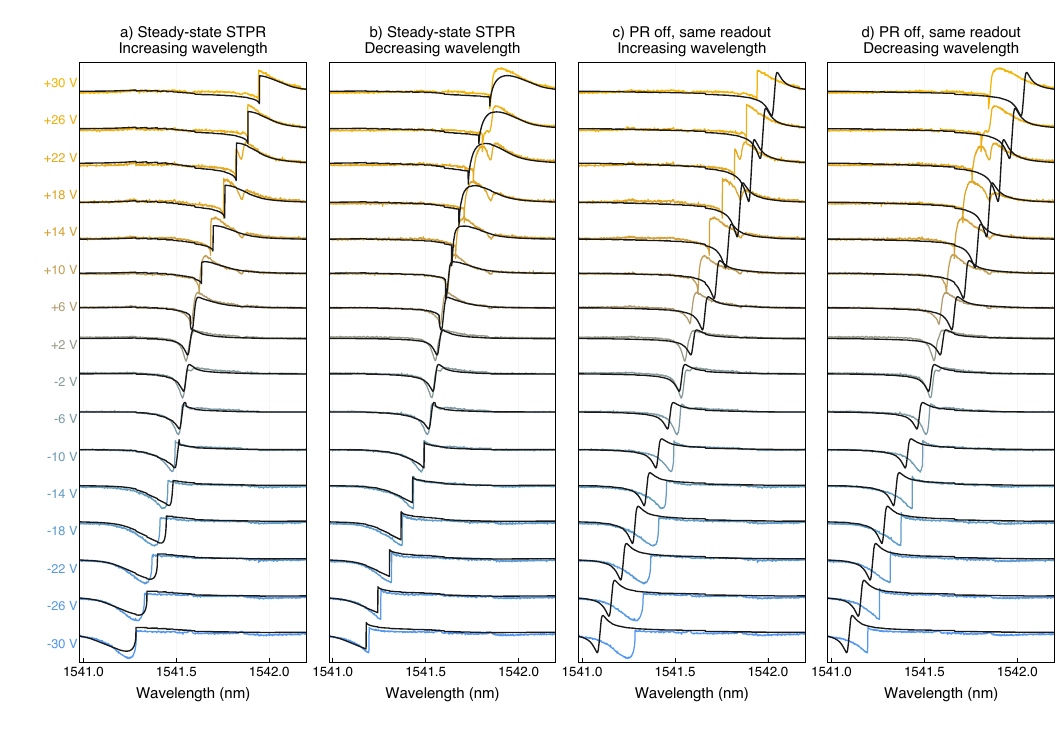}
\caption{Measured spectra (colored by applied voltage) and modeled
  spectra (black) for cavity $C_3$ at $2\,\mathrm{nm\,s^{-1}}$ and an
  estimated cavity-arm input power of $-11.8\,\mathrm{dBm}$. (a,b)~STPR model combined with the three-arm
  interferometer model for increasing- and
  decreasing-wavelength sweeps; (c,d)~control with zero photorefractive shift and the
  same readout and gain as Fig.~\ref{fig:C3spectra}. Median normalized
  trace errors in the supplementary scoring windows are $12.4\%$ and
  $39.6\%$, respectively.}
\label{fig:wf_C3}
\end{figure}

\begin{figure}[t]
\centering
\includegraphics[width=\linewidth]{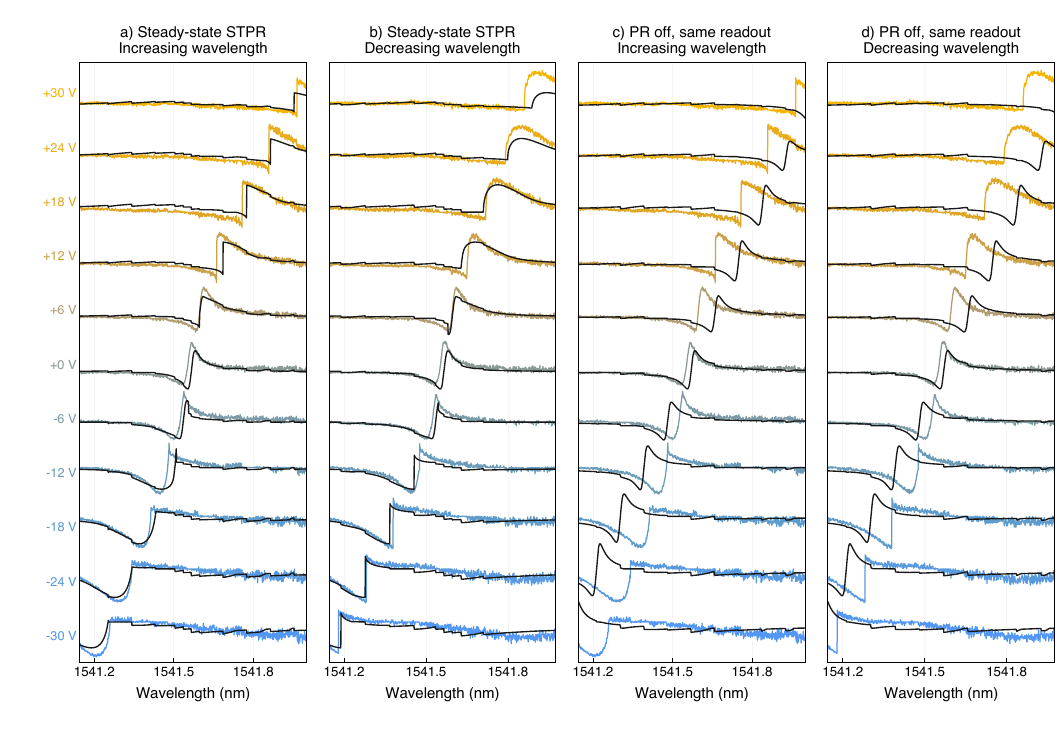}
\caption{Measured spectra of the acquisition shown in
  Fig.~\ref{fig:device}(c) (cavity $C_1$, $1\,\mathrm{nm\,s^{-1}}$,
  estimated cavity-arm input power $-16.8\,\mathrm{dBm}$, all $11$ voltage steps),
  with the fixed effective $3\times3$ readout. Colored curves are measured;
  black curves are modeled. (a,b)~Quasi-static STPR for increasing- and
  decreasing-wavelength sweeps; (c,d)~control with zero photorefractive shift and the
  same optical readout and gain. Up and down sweeps at each voltage use the
  same measured-percentile display scale; neither model curve is clipped or
  shifted. The down-sweep panels omit the scan-start discontinuity above
  $1541.975\,\mathrm{nm}$; the omitted samples are also excluded from the
  error metric. Median normalized errors are $12.5\%$ and $27.4\%$,
  respectively.}
\label{fig:wf_fig1c}
\end{figure}

Supplementary Figs.~\ref{fig:wf_C3} and~\ref{fig:wf_fig1c} compare
measured spectra with the standard STPR pole trajectories passed through an
effective reciprocal $3\times3$ interferometric readout. The cavity-specific
loaded $Q$ values are estimated from Fano fits to the measured resonances;
the waterfall readout itself has no fitted Fano line shape. The physical STPR
parameters are fixed by the resonance-position fits. The
Fig.~\ref{fig:device}(c) acquisition retains its archived low-power
wavelength calibration, as specified below. The optical parameters and detector-port assignment are shared
across voltages, sweep directions, and outputs within each acquisition.
For the control, we set $\Vscmax=0$ to remove the photorefractive resonance
shift while retaining the same optical readout and displayed gain.
The figures label this control \emph{PR off}. These are calibrated, effective isolated-cavity
readouts, not independently calibrated predictions of absolute three-port
transmission.

Following the transfer-matrix description of the $3\times3$
switch~\cite{BuzziRouter2026}, we denote the forward MMI transfer matrix by
$\mathbf M$ and the effective arm reflections by
$\mathbf R=\operatorname{diag}(r_1,r_2,r_3)$. Reciprocity gives the
return-pass matrix $\mathbf M^{\mathsf T}$, so the round-trip response is
\begin{equation}
  \mathbf T=\mathbf M^{\mathsf T}\mathbf R\mathbf M,
  \qquad \mathbf E_{\mathrm{out}}=\mathbf T\mathbf e_0,
  \qquad P_j=G_j|E_{\mathrm{out},j}|^2.
  \label{eq:routerreadout}
\end{equation}
We use the ideal equal-split convention
$M_{jk}=3^{-1/2}\exp(2\pi i jk/3)$ for $j,k\in\{0,1,2\}$, with input at port $0$.
The resonant arm has response
$r_c=\exp(i\beta\Vapp)\left[1-2\eta/(1+i(\lambda-\lambda_c)/\gamma)\right]$,
where $\lambda_c$ is the instantaneous cavity resonance and
$\gamma=\lambda_0/(2Q)$. The other two arms have effective responses
$a_m\exp\{i[\phi_m+s_m(\lambda-\lambda_*)]\}$.
The port assignment is fixed within each acquisition. Coherent mixing in
Eq.~\eqref{eq:routerreadout} generates the asymmetric spectral features
without a separate Fano fit at every voltage.

For the resonance-position fits, we calibrate the existing eight
optical coefficients ($\eta$, two reference-arm amplitudes, two phases,
two wavelength slopes, and $\beta$) separately for the original $C_1$,
$C_2$, and $C_3$ acquisitions and the later $C_1$ acquisition. The detector
indices are 0, 2, 1, and 1, respectively (zero-based indexing); their optical
port assignments are fixed. Calibration uses the lowest-power spectra at
$2\,\mathrm{nm\,s^{-1}}$ for the original acquisitions and
$1\,\mathrm{nm\,s^{-1}}$ for the later acquisition, with alternating
voltage rows used for fitting. A preliminary steady-state trajectory is
held fixed while estimating the readout. The full intensity from
Eq.~\eqref{eq:routerreadout} is evaluated before applying the identical
40th-percentile background subtraction to measured and simulated sequences.
We do not remove a cavity-free model response before tracking.

The calibrated readout is then held fixed during all resonance-position
fits and residual resamples. Its coefficients describe an effective
interferometric response, not uniquely measured arm properties; some
coefficients reach their calibration bounds. The conditional parameter
ranges do not include uncertainty in this optical calibration. Background
estimation uses all voltage rows, so interleaved-row spectral checks are
conditional on that common preprocessing.

Figure~\ref{fig:C3spectra} uses the $C_3$ parameters in
Table~\ref{tab:perchannel}, its fixed optical coefficients, and one positive
gain calibrated on alternating voltage rows at the displayed power.
Measured and modeled display traces receive the same 40th-percentile
background subtraction and share a vertical normalization at each voltage.
Each trace error is its RMS residual over
$1540.98$--$1542.20\,\mathrm{nm}$, normalized by the measured
up/down pair's 1st--99th percentile span. The median normalized trace
errors are $9.4\%$ on the calibration rows
and $10.3\%$ on the interleaved rows;
the RMS across all 32 trace errors is $10.3\%$.

Supplementary Fig.~\ref{fig:wf_C3} uses the same physical parameters,
reference-arm coefficients, detector assignment and positive gain as
Fig.~\ref{fig:C3spectra}. The zero-shift control retains that calibration
and gain.
The same 40th-percentile background is subtracted from each sequence, with no
individually fitted trace gain, offset, phase or wavelength shift.
Errors use a voltage-dependent window within $\pm0.3\,\mathrm{nm}$
of the archived electro-optic tuning line, whereas Fig.~\ref{fig:C3spectra}
uses the fixed $1540.98$--$1542.20\,\mathrm{nm}$ window. Both use the
measured up/down pair's $1$st--$99$th percentile span for normalization.
The different windows give different error statistics for the same spectra.
For Fig.~\ref{fig:wf_C3}, the median over 32 traces is $12.4\%$
for steady-state STPR and $39.6\%$ for the control; RMS across traces
is $13.1\%$ and $40.6\%$, respectively.

For Supplementary Fig.~\ref{fig:wf_fig1c}, the physical $C_1$
parameters are fixed at the values used in Fig.~\ref{fig:K1}. We fit
$\rV$ and $\lambda_0$ to this acquisition's extracted resonance wavelengths
at the two lowest input powers, using the same equal-power wavelength
cost function. This gives $\rV=14.89\,\mathrm{pm\,V^{-1}}$ and
$\lambda_0=1541.5710\,\mathrm{nm}$. The eight interferometer coefficients,
determined from the two lowest-power spectra using alternating voltages,
both scan directions, and detector 1, remain fixed.
One positive gain per calibration power is shared across its voltages
and directions. At the displayed power, one positive gain is fitted on
alternating voltages and held fixed for both the STPR and zero-shift
predictions. This is a calibrated spectral comparison rather than an
entirely held-out power test. Both measured and modeled full intensities
receive the same 40th-percentile background subtraction, with no individual
trace gain, phase, offset, or wavelength shift.

The initial $1541.975$--$1542.0\,\mathrm{nm}$ portion of decreasing
sweeps is excluded because of a reproducible scan-start discontinuity;
increasing sweeps retain $1541.145$--$1542.0\,\mathrm{nm}$.
Each trace error is the RMS difference over displayed samples normalized
by the measured up/down pair's 1st--99th percentile span. The median over
22 traces is $12.5\%$ for STPR and
$27.4\%$ with the photorefractive shift set to zero.
The scoring window differs from Supplementary Fig.~\ref{fig:wf_C3}.

\end{document}